\documentclass[a4paper,10pt]{article}
\usepackage[utf8]{inputenc}
\usepackage[english]{babel}
\usepackage{etoolbox}
\apptocmd{\sloppy}{\hbadness 10000\relax}{}{}

\usepackage{comment}

\usepackage[usenames]{color}

\usepackage{dcolumn,booktabs}
\newcommand\mc[1]{\multicolumn{1}{c}{\text{#1}}} % handy shortcut macro
\newcolumntype{d}{D{.}{.}{-1}} %from apsrev4-2

\usepackage{jheppub}

\hypersetup{unicode=true,pdfauthor={Edgardo Franzin, Antonia M. Frassino, and Jorge V. Rocha},bookmarksnumbered=false,bookmarksopen=false, breaklinks=true,colorlinks=true,citecolor=blue}
\makeatletter\g@addto@macro\bfseries{\boldmath}\makeatother%

\def\be#1\ee{\begin{align}#1\end{align}}
\def\nn{\nonumber}

\def\gsim{\, \rlap{$>$}{\lower 1.1ex\hbox{$\sim$}}\,}
\def\lsim{\, \rlap{$<$}{\lower 1.1ex\hbox{$\sim$}}\,}

\DeclareFontFamily{U}{mathx}{\hyphenchar\font45}
\DeclareFontShape{U}{mathx}{m}{n}{<-> mathx10}{}
\DeclareSymbolFont{mathx}{U}{mathx}{m}{n}
\DeclareMathAccent{\widebar}{0}{mathx}{"73}
\def\ol{\widebar}

\allowdisplaybreaks

\makeatletter
\renewcommand{\@fnsymbol}[1]{\ifcase#1\or a\or b\or c\or d\or e\or f\or g\or h\or i\or j\or k\else\@ctrerr\fi}
\makeatother

\definecolor{purple}{rgb}{0.7,0,1}

\definecolor{green}{rgb}{0,0.7,0.2}

\title{Tidal Love numbers of static black holes in anti-de Sitter}

\author[a,b]{Edgardo Franzin}%\,\orcidlink{0000-0002-5705-5550}}
\author[c,d,e,f]{Antonia M.~Frassino}%\,\orcidlink{0000-0001-9937-486X}}
\author[g,h,i]{Jorge V.~Rocha}%\,\orcidlink{0000-0001-9619-7087}\,}

\affiliation[a]{Dipartimento di Fisica, ``Sapienza'' Universit\`a di Roma, p.le Aldo Moro~5, 00185, Rome, Italy}
\affiliation[b]{INFN, Sezione di Roma1, p.le Aldo Moro~2, 00185, Rome, Italy}

\affiliation[c]{SISSA, International School for Advanced Studies, via Bonomea 265, 34136 Trieste, Italy}
\affiliation[d]{INFN, Sezione di Trieste, via Valerio 2, 34127 Trieste, Italy}
\affiliation[e]{Departamento de F\'{i}sica y Matem\'{a}ticas, University of Alcal\'{a}, Campus universitario 28805, Alcal\'a de Henares (Madrid), Spain} 
\affiliation[f]{Departament de F\'isica Qu\`antica i Astrof\'isica, Institut de Ci\`ences del Cosmos, Universitat de Barcelona,
Mart\'i i Franqu\`es 1, 08028 Barcelona, Spain}

\affiliation[g]{Departamento de Matem\'atica, ISCTE--Instituto Universit\'ario de Lisboa, Avenida das For\c{c}as Armadas, 1649-026 Lisboa, Portugal}
\affiliation[h]{Centro de Astrof\'isica e Gravita\c{c}\~ao--CENTRA, Instituto Superior T\'ecnico--IST, Universidade de Lisboa--UL, Av.\ Rovisco Pais 1, 1049-001 Lisboa, Portugal}
\affiliation[i]{Instituto de Telecomunica\c{c}\~oes--IUL, Avenida das For\c{c}as Armadas, 1649-026 Lisboa, Portugal}

\emailAdd{edgardo.franzin@uniroma1.it}
\emailAdd{afrassin@sissa.it}
\emailAdd{jorge.miguel.rocha@iscte-iul.pt}

\abstract{Tidal Love numbers of anti-de Sitter black holes are understood as linear response  coefficients governing how the holographically dual plasma polarizes when the geometry of the space, in which the plasma lives, is deformed. So far, this picture has been applied only to black branes with plane wave perturbations. We fill the gap in the literature by performing the computation of tidal Love numbers for the four-dimensional Schwarzschild solution in global anti-de Sitter, which is dual to a conformal plasma on $S^2$. We conclude about the effect of the bulk gravitational perturbations on the boundary metric and stress tensor, responsible for the geometric polarization.
The computation of the tidal Love numbers is performed in both Regge-Wheeler gauge and the Kodama-Ishibashi gauge-invariant approach. We spell out how to convert the tidal Love numbers determined in these two formalisms and find perfect agreement. We also relate the Kodama-Ishibashi formalism with the Kovtun-Starinets approach, which is particularly well suited for the holographic analysis of black branes. This allows us to compare with the tidal Love number results for black branes in anti-de Sitter, also finding agreement in the relevant regime.}

\begin{document}

\maketitle

% numbered footnotes
%\renewcommand*{\thefootnote}{\arabic{footnote}}

%\tableofcontents

%%%%%%%%%%%%%%%%%%%%%%%%%%%
%%%%%%%  SECTION 1  %%%%%%%
%%%%%%%%%%%%%%%%%%%%%%%%%%%
\section{Introduction
\label{s:Intro}}
%%%%%%%%%%%%%%%%%%%%%%%%%%%

Tidal Love numbers (TLNs) characterize the deformability of a self-gravitating object~\cite{Binnington:2009bb,Damour:2009vw}. More precisely, they are defined as the (real part of the) linear response of the mass and current multipole moments of the object to applied external tidal fields. Consequently, they are consistently (and traditionally) computed within the framework of linear gravitational perturbation theory~\cite{Fang:2005qq, Flanagan:2007ix, Damour:2009vw, Binnington:2009bb}, although more recently non-linear TLNs have also been under scrutiny~\cite{DeLuca:2023mio, Riva:2023rcm, Combaluzier-Szteinsznaider:2024sgb, Kehagias:2024rtz}.
TLNs obviously depend on the internal structure of the self-gravitating body under consideration, but also on the gravitational field dynamics. For the case of vacuum black holes (BHs) ---the type of object we study exclusively in this article--- only the gravito-static equations of the gravity theory are relevant.
Tidal environments for self-gravitating bodies occur naturally in compact binaries, and therefore TLNs are relevant for gravitational-wave astronomy. Indeed, tidal coupling between compact objects during the late (inspiral) stage of a binary merger affects the gravitational waveform~\cite{Cardoso:2017cfl, Flanagan:2007ix, Hinderer:2016eia}.

It is known that stationary, asymptotically flat, vacuum BHs in general relativity (GR), concretely in four spacetime dimensions, have vanishing TLNs~\cite{Binnington:2009bb, Damour:2009vw, Kol:2011vg, Gurlebeck:2015xpa, LeTiec:2020spy, LeTiec:2020bos, Chia:2020yla}.
This perplexing result is now understood as a direct consequence of hidden (or ladder) symmetries recently uncovered~\cite{Charalambous:2021kcz, Hui:2021vcv}. 
Generically, similar symmetry structure (and the resulting vanishing of the TLNs of four-dimensional BHs in GR) does not carry over to the non-asymptotically flat BHs arising as vacuum solutions of the Einstein equations with the inclusion of a non-vanishing cosmological constant. 

The aim of the present paper is to investigate the effect of a negative cosmological constant on the TLNs of spherically symmetric BHs. In particular, we compute the TLNs of (non-rotating) four-dimensional asymptotically anti-de Sitter (AdS) BHs. As expected, according to the discussion above, they show non-trivial behavior.
We restrict our study to Schwarzschild-AdS BHs corresponding to negative values of the cosmological constant.  The de Sitter analysis has been recently presented in~\cite{Nair:2024mya} using, however, a different approach, namely the wordline effective field theory approach. The study of TLNs of BH spacetimes is rapidly growing, with recent developments in several directions --- see for example~\cite{Capuano:2024qhv, Rai:2024lho, DeLuca:2024ufn, Atkins:2023axs, Katagiri:2022vyz, Katagiri:2023umb, Katagiri:2024fpn, Arana:2024kaz}.

The asymptotically AdS setup is compelling, with TLNs of AdS BHs acquiring a solid holographic interpretation via the celebrated AdS/CFT correspondence~\cite{Maldacena:1997re,Gubser:1998bc,Witten:1998qj}. As pointed out in~\cite{Emparan:2017qxd}, the TLNs encode the geometric polarizability of the conformal plasma dual to the BH.
Moreover, the physics of BHs in AdS spacetime is interesting in its own right, with a rich phase diagram~\cite{Hawking:1982dh, Chamblin:1999tk, Marolf:2013ioa}. Related to this, one should bear in mind that the so-called small BHs in AdS are not thermodynamically stable, they do not have such a clear holographic interpretation~\cite{Horowitz:1999jd}, so our results for the geometric polarizability of the associated plasma in those cases should be taken with a grain of salt.
At the level of the holographic dual description, the boundary conformal field theory (CFT) for global AdS BHs lives on a sphere, whereas in~\cite{Emparan:2017qxd} the dual plasma lives on a plane. In the latter (former) context, the TLNs control how the stress tensor of the plasma adjusts to deformations of the plane (sphere). Our results address the effect of background curvature of the boundary of AdS on the TLNs. 

Curiously, TLNs have not been computed for BHs in global AdS so far, so one of the aims of this paper is to fill this gap in the literature. TLNs have been determined explicitly for AdS black \textit{branes}~\cite{Emparan:2017qxd}, whose metric can be recovered exactly as the large mass limit of the Schwarzschild-AdS BH, i.e., by taking the Schwarzschild radius to be much larger than the AdS length scale. Linearized perturbations of the BH with large spherical harmonic multipole number $l$ (compared to the BH mass in AdS units) can be similarly mapped into perturbations of the black brane. Our results are expected to approach those of~\cite{Emparan:2017qxd} in this limit, and this is indeed the case.
However, the comparison is not as straightforward to make as one might think, being overshadowed by technical details, which we now overview. 

Ref.~\cite{Emparan:2017qxd} adopted the Kovtun-Starinets (KS) gauge-invariant formalism~\cite{Kovtun:2005ev} to solve the gravitational perturbation equations, which is especially well-suited to deal with black-brane geometries. In addition, having a background metric that is translationally invariant along the field theory directions allowed the decomposition of the (static) perturbations in plane waves. Those modes are defined by their wavenumber, which is a parameter that takes values in $\mathbb{R}$.
In contrast, linear perturbations of static, spherically symmetric BHs are naturally decomposed in spherical harmonics, giving rise to a \emph{discrete} set of modes. Moreover, most of the literature dealing with linear perturbations of spherically symmetric BHs adopt the Regge-Wheeler (RW) gauge~\cite{Regge:1957td,Zerilli:1970se}, and it is not immediate how to relate quantities (in our case, we are interested in the TLNs) computed in that gauge with equivalent quantities computed in the KS formalism.
We will circumvent this difficulty by resorting to a third approach to tackle BH linear perturbations:\ the well-known Kodama-Ishibashi (KI) gauge-invariant formalism~\cite{Kodama:2003jz,Ishibashi:2011ws}. The advantage of adopting the KI approach is that it is equally suitable for BHs \emph{and} for black branes, in addition to being a gauge-invariant scheme. It will serve as the link allowing us to connect the RW and KS formalisms. 

At any rate, whatever formalism is employed, the equations governing linear gravitational perturbations are of second order, since they descend from the cosmological Einstein equations (also of second order). In all three formalisms, the full set of linearized equations boil down to just two decoupled master equations, each of which generically admits two linearly independent solutions.
For asymptotically flat backgrounds, typically one of these modes grows with the radial coordinate, while the other decays (in the RW approach). This, however, does not extend to geometries with AdS asymptotics, where the two possible modes are both decaying in the polar sector. The relevant distinction between the two modes is whether they are, or not, \emph{normalizable}. As we will see, identifying which one is (non-)normalizable can be tricky when working in the RW gauge-fixed approach. Also in this respect it is useful to map all our results to a gauge-invariant formalism.

Once the non-normalizable and normalizable modes are identified, any solution of the master (linear) perturbation equation for a given sector is uniquely written as a linear combination of the former.
TLNs are then straightforwardly defined as dimensionless ratios of the coefficients of the normalizable and non-normalizable terms for regular perturbations. For each perturbation type ---defined by the sector (polar or axial) and by the usual spherical harmonics indices $(l,m)$--- one may therefore compute an associated TLN.
This can be done within any formalism chosen to address the gravitational perturbations, and we will compute the TLNs in both RW and KI approaches. The mapping between the respective master equations mixes them non-trivially. As a result the TLN of a given perturbation type computed in the RW gauge is generically distinct from the TLN of the same mode computed in the KI formalism. Nevertheless, we are able to provide explicit formulas to convert between the TLNs computed in the two approaches. 

The preceding paragraph may raise some concerns about the physical significance of the TLNs we compute in AdS. However, as already mentioned, for asymptotically AdS BHs they represent the geometric polarizability of the holographically dual plasma~\cite{Emparan:2017qxd}. Therefore, apart from the well-understood dependencies on the holographic renormalization scheme, the physical TLNs have a well-defined meaning in terms of the boundary stress tensor, where they may be defined by the coefficients of the linear perturbations induced by the tidal forcing terms, i.e., they control the linear response of the dual plasma system. This interpretation is analogous to the invariant meaning of TLNs in asymptotically flat spacetimes where they are regarded as Wilson coefficients in the worldline effective field theory~\cite{Goldberger:2004jt, Porto:2016pyg, Charalambous:2021mea}.

With the exception of pure AdS (i.e., no BH present, or in the related eikonal limit) the master perturbation equations in the Schwarzschild-AdS background do not lend themselves to analytic treatment and must be solved numerically. Accordingly, we obtain the TLNs of static AdS BHs numerically in each of the two approaches (RW and KI), finding precise agreement upon using the conversion formulas mentioned above. 

It is important to keep in mind that the asymptotically flat limit is not obtained continuously from our results simply by sending the cosmological constant to zero. As alluded to above, the TLNs of a Schwarzschild BH exactly vanish, and this is not obtained directly as a limit of the Schwarzschild-AdS TLNs.
The origin of this discontinuity will be explained in Secs.~\ref{s:LoveSAdS4} and~\ref{s:LoveKI}.
The fact that static BHs in vacuum GR have vanishing TLNs is a well-known result that has been demonstrated adopting the RW gauge~\cite{Binnington:2009bb,Damour:2009vw,Fang:2005qq}, but in Appendix~\ref{a:LoveKISchwarz} we provide an equivalent demonstration using the KI approach.

The rest of the paper is organized as follows.
In Sec.~\ref{s:Love_gen}, we specify the BH spacetimes we shall restrict to, describe the general formalism to assess their gravitational perturbations adopting the RW gauge, and present the corresponding master equations.
Section~\ref{s:LoveSAdS4} describes the computation of the TLNs for the Schwarzschild-AdS BH in the RW approach. 
In Sec.~\ref{s:LoveKI} we recall the gauge-invariant KI master equations and obtain the TLNs in this formalism, relating them to the results of Sec.~\ref{s:LoveSAdS4}.
Section~\ref{s:holoLove} evokes the AdS/CFT correspondence, and in particular the holographic renormalization approach, to compute the boundary stress tensor and metric corresponding to tidally perturbed AdS BHs.
Finally, we draw our conclusions and present a few speculations in Sec.~\ref{s:discussion}.
We relegate several technical computations to the appendices. This includes the comparison of our results to those obtained for black branes in the appropriate limit.

Throughout this paper, we use units $G=c=1$.

%%%%%%%%%%%%%%%%%%%%%%%%%%%
%%%%%%%  SECTION 2  %%%%%%%
%%%%%%%%%%%%%%%%%%%%%%%%%%%
\section{Perturbation equations for spherically symmetric black holes\label{s:Love_gen}}
%%%%%%%%%%%%%%%%%%%%%%%%%%%

Our focus will be exclusively on spherically symmetric, static background solutions, for which the metric can be expressed as
\be
\label{eq:metric_original}
ds^2 = g_{\mu\nu}dx^\mu dx^\nu = - e^{\psi(r)}\,dt^2 + e^{\lambda(r)}\,dr^2 + r^2\,d\Omega_2^2,
\ee
being $d\Omega_2^2 \equiv d\theta^2 + \sin^2\theta\,d\phi^2$, and we are interested in \emph{static} perturbations of a similar background solution $\ol{g}_{\mu\nu}$.
As usual, the perturbed metric is written as
$g^{\,\text{pert}}_{\mu\nu} = \ol{g}_{\mu\nu} + \delta g_{\mu\nu} = \ol{g}_{\mu\nu} + h_{\mu\nu}\,$, where
$h_{\mu\nu} \ll \ol{g}_{\mu\nu}\,$. 
Adopting the RW gauge~\cite{Regge:1957td}, the metric perturbation is separated according to parity into polar and axial sectors, $h_{\mu\nu}=h_{\mu\nu}^\text{polar}+h_{\mu\nu}^\text{axial}$, each of which is further decomposed into spherical harmonics,
\be
\label{heven}
h_{\mu\nu}^\text{polar} &=
\begin{pmatrix}
-e^{\ol{\psi}} H_0^{lm}\!\! & H_1^{lm} & 0 & 0 \\
 H_1^{lm} & e^{\ol{\lambda}} H_2^{lm}\!\! & 0 &0\\
 0 &0 & r^2 K^{lm}\!\! & 0\\ 
0&0 &0 & \!\! r^2\sin^2{\theta} K^{lm}\!\!
\end{pmatrix} Y^{lm},\\
\label{hodd}
h_{\mu\nu}^\text{axial} &=
\begin{pmatrix}
0 & 0 & h_0^{lm}S_\theta^{lm} & h_0^{lm}S_\phi^{lm}\;\; \\
0 &0 & h_1^{lm}S_\theta^{lm} & h_1^{lm}S_\phi^{lm} \;\;\\
\;\;h_0^{lm}S_\theta^{lm} &h_1^{lm}S_\theta^{lm}  &0 &0\\
\;\;h_0^{lm}S_\phi^{lm}& h_1^{lm}S_\phi^{lm} & 0 &0
\end{pmatrix},
\ee
where $Y^{lm}(\theta,\phi)$ stand for the scalar spherical harmonics and
$S_\theta^{lm}\equiv -Y_{,\phi}^{lm}/\sin{\theta}$, $S_\phi^{lm}\equiv\,\sin{\theta} \,Y_{,\theta}^{lm}$
are obtained from their partial derivatives with respect to $\phi$ and $\theta$. The functions $H_0^{lm}(r)$, $H_1^{lm}(r)$, $H_2^{lm}(r)$ and $K^{lm}(r)$ are yet undetermined functions of $r$ in the polar sector. Similarly, $h_0^{lm}(r)$ and $h_1^{lm}(r)$ are undetermined functions of $r$ in the axial sector. The bar above $\psi(r)$ and $\lambda(r)$ indicates those metric functions are to be evaluated on the specific background considered.

Since the background is spherically symmetric, the perturbation equations cannot mix terms with different parity (polar or axial) nor different $l$; moreover, as the master equations do not depend on $m$, this can be set equal to zero for the purpose of deriving the perturbation equations. Therefore, in the following we drop both indices $l$ and $m$ from the metric functions. 
We will reinstate the index $l$ when we present the formulas for the TLNs, to stress that those quantities are $l$-dependent.
Notice that some of the expressions obtained below assume $l\neq1$, but this does not represent a limitation since we justify in Sec.~\ref{s:LoveSAdS4} that the only gravitational perturbation modes of concern to us have $l\geq2$. 

We take the background metric to be an Einstein mani\-fold, thus satisfying $\ol{R}_{\mu\nu} = \Lambda \ol{g}_{\mu\nu}$. 
Birkhoff's theorem then guarantees the unperturbed spacetime is described by the line element~\eqref{eq:metric_original} with the non-trivial metric functions given by
\be
e^{\ol{\psi}(r)} = e^{-\ol{\lambda}(r)} = 1 - \frac{2M}{r} - \frac{\Lambda r^2}{3}\,,
\label{eq:bkgd_metric}
\ee
where $M$ and $\Lambda$ are the mass of the BH and the cosmological constant, respectively. This corresponds to the four-dimensional Schwarzschild-AdS BH solution.
The single positive real root of $e^{-\ol{\lambda}}$ yields the radial location of the event horizon. We shall denote it by $r_h$ and refer back to this only in Sec.~\ref{s:LoveSAdS4}.

Assuming merely that the background has constant  curvature one gets the standard linearized perturbation equations in GR. In this case, it is well known that one may impose, without loss of generality, the transverse-traceless gauge conditions, $\ol{\nabla}^\alpha h_{\mu\alpha}=0\,, \, h=\ol{g}^{\mu\nu}h_{\mu\nu}=0$ and get
\be
\label{fofRperts}
  -{\ol{R}^{\alpha}}{}_\mu {{}^\beta}{}_\nu \, h_{\alpha\beta}
  - \frac{1}{2} \ol{\nabla}^2 h_{\mu\nu} = 0.
\ee

Let us focus first on the polar sector. Inserting a metric perturbation of the form~\eqref{heven} in~\eqref{fofRperts}, one finds that $H_1(r)$ vanishes while $H_2 = -H_0$.
Denoting $H(r) \equiv H_0(r)$, the remaining free function $K(r)$ is completely fixed by
\be
K(r) = \frac{e^{-\ol{\lambda}(r)}}{(l+2)(l-1)} \left\{ \left[ 2 - l(l+1) e^{\ol{\lambda}(r)} - r^2 \ol{\psi}'(r)^2 \right] H(r) - r^2 \ol{\psi}'(r) H'(r) \right\}\,.
\label{eq:Kconstraint}
\ee
The polar metric perturbations are then governed by a single equation,
\be
H'' + \frac{1}{2} \Big\{ \frac{4}{r} + \ol{\psi}' - \ol{\lambda}' \Big\}H' 
+ \frac{1}{2r^2} \Big\{ r \Big[ \ol{\lambda}' \left(2 - r\ol{\psi}'\right) + 2r\ol{\psi}'' - r\ol{\psi}'^2 + 6\ol{\psi}' \Big] - 2e^{\ol{\lambda}} l (l+1) \Big\}H = 0\,,
\label{eq:Hpertgen}
\ee
where the prime denotes a derivative with respect to $r$.
By requiring that the background metric satisfies the Einstein equations with a cosmological constant, one obtains
\be
\ol{\lambda}'(r) &= -\ol{\psi}'(r) = \frac{1 - e^{\ol{\lambda}(r)} \left(1 - \Lambda r^2\right)}{r}\,, \\
\ol{\psi}''(r) &= \frac{1 -2e^{\ol{\lambda}(r)} \Lambda r^2 - e^{2 \ol{\lambda}(r)} \left(1 - \Lambda r^2\right)^2}{r^2}\,.
\ee
Replacing these equalities in~\eqref{eq:Hpertgen} yields the polar sector equation in a form that depends explicitly only on the background function $\ol{\lambda}(r)$, but not on any of its derivatives,
\be
H'' + \frac{1}{r}\left\{ 1 + e^{\ol{\lambda}}\left(1-\Lambda r^2\right) \right\}H' 
- \frac{1}{r^2} \left\{ 1 + e^{\ol{\lambda}} \Big[(l+2)(l-1) + 4\Lambda r^2 \Big] + e^{2\ol{\lambda}} \left(1-\Lambda r^2\right)^2 \right\}H = 0\,. \label{eq:Hpert}
\ee

We now move on to the axial sector. It turns out that linear perturbations of this type automatically satisfy $\ol{\nabla}^\alpha \ol{\nabla}^\beta h_{\alpha\beta} = 0 = h$.
The equations of motion further require $h_1(r)=0$ and the perturbation equations for this sector reduce to 
\be
h_0'' - \frac{\ol{\lambda}'+\ol{\psi}'}{2} h_0' - \frac{2 + (l+2)(l-1) e^{\ol{\lambda}} - r \left(\ol{\lambda}' + \ol{\psi}'\right) }{r^2} h_0 = 0\,.
\ee
Again imposing the replacements for the derivatives of the background functions, the previous equation becomes
\be
\label{eq:h0pert}
h_0'' - \frac{1}{r^2} \left[ 2 + e^{\ol{\lambda}}(l+2)(l-1) \right] h_0 = 0\,.
\ee

Note that, differently from the polar sector, this equation only depends on the cosmological constant $\Lambda$ \textit{implicitly} through the background function $\ol{\lambda}(r)$.

%%%%%%%%%%%%%%%%%%%%%%%%%%%
%%%%%%%  SECTION 3  %%%%%%%
%%%%%%%%%%%%%%%%%%%%%%%%%%%
\section{Love numbers of Schwarzschild-AdS\texorpdfstring{$_4$}{4} in the Regge-Wheeler formalism\label{s:LoveSAdS4}}
%%%%%%%%%%%%%%%%%%%%%%%%%%%

Now that we have the general equations governing polar and axial static perturbations of spherically symmetric solutions of the Einstein equations with a cosmological constant, we particularize to the four-dimensional Schwarzschild-AdS BH.

Evaluated on the background defined by~\eqref{eq:bkgd_metric}, the perturbation equations~\eqref{eq:Hpert} and \eqref{eq:h0pert} are written explicitly as
\be
H'' &+ \frac{4\Lambda r^3 - 6r + 6M}{r \left(\Lambda r^3 - 3r + 6M\right)}H'\nn\\
&+ \frac{2\Lambda^2 r^6 +3(l-2)(l+3)\Lambda r^4 + 60M\Lambda r^3 - 9l(l+1)r^2 + 18l(l+1)Mr - 36M^2}{r^2 \left(\Lambda r^3 - 3r + 6M\right)^2}\,H = 0\,,\label{eq:Polar-AdS}
\ee
for the polar sector, and
\be
h_0'' - \frac{2\Lambda r^3 -3 l(l+1) r + 12M}{r^2 \left(\Lambda r^3 - 3r + 6M\right)}\,h_0 = 0\,,\label{eq:Axial-AdS}
\ee
for the axial sector. 

Equations~\eqref{eq:Polar-AdS} and~\eqref{eq:Axial-AdS} depend explicitly on two dimensionful quantities, namely the mass $M$ and the cosmological constant $\Lambda$. The latter defines a length scale ---the AdS scale--- through
\be
L\equiv\sqrt{-3/\Lambda}\,.
\ee
Accordingly, we implicitly assume that the cosmological constant is strictly negative, as is appropriate for AdS spacetime. By expressing the radial variable and the mass in AdS units, as
\be
\rho\equiv \frac{r}{L}\,,  \qquad
\mu \equiv \frac{M}{L}\,,
\ee
$L$ cancels out in both~\eqref{eq:Polar-AdS} and~\eqref{eq:Axial-AdS}, which are converted into differential equations in $\rho$ depending on a single dimensionless parameter, $\mu$.

We now proceed to evaluate the TLNs of the Schwarzschild-AdS$_4$ BH for each of the above two sectors separately. Naturally, these quantities characterizing the tidal deformability of the object will depend, apart from $\mu$, on the multipole number $l$.
Similarly to what occurs with a Schwarzschild background, the $l=0$ and $l=1$ modes should be dismissed in the study of (static) gravitational perturbations of the Schwarzschild-AdS BH. The analysis of~\cite{Regge:1957td,Zerilli:1970se} that justifies this in the asymptotically flat case goes through with minor adjustments when a negative cosmological constant is included.
The polar $l=0$ mode just induces a change of the mass, in accordance with Birkhoff's theorem, and the axial $l=1$ mode adds angular momentum to the background BH. These quantities must be conserved in spite of tidal effects. 
On the other hand, the polar $l=1$ mode generates a shift in the center of mass that can be completely eliminated by a gauge transformation, while the axial $l=0$ mode vanishes identically.
The upshot is that (static) TLNs of Schwarzschild-AdS only exist for $l\geq2$.

%%%%%%%%%%%%%%%%%%%%%%%%%%%
\subsection{Polar sector\label{s:LovePolar}}

Asymptotically, the solutions to Eq.~\eqref{eq:Polar-AdS} are linear combinations of a normalizable and a non-normalizable function,
\be
H \sim \frac{C_-^\text{p}}{r} \left[1 - \frac{3(l^2+l-4)}{2\Lambda r^2} + \dots \right] + \frac{C_+^\text{p}}{r^2} \left[ 1 - \frac{(l-2)(l+3)}{2\Lambda r^2} + \dots \right]\,,
\label{HinfRW}
\ee
and we define the dimensionless polar TLNs as
\be\label{polarLoveRW}
k_l^\text{polar} \equiv L\,\frac{C_-^\text{p}}{C_+^\text{p}}\,.
\ee
Note that it is the coefficient $C^{\rm p}_+$ that controls the non-normalizable term, even though it multiplies the term in Eq.~\eqref{HinfRW} with the fastest decay $\sim r^{-2}$. The reason for this counter-intuitive association is that a non-trivial $H(r)$ implies also a non-vanishing $K(r)$ in the polar perturbations, and it turns out that $C^{\rm p}_+$ is the coefficient of the term in $K(r)$ which grows like $r^2$, i.e., it governs the non-normalizable term. 

Our task is now to numerically integrate~\eqref{eq:Polar-AdS} for various values of the multipole number $l$ and of the quantity $\mu$.
We implemented two numerical routines to compute the TLNs, which we now briefly describe.
Regular boundary conditions at the event horizon require $H(r) = O(r-r_h)$ while at infinity the function $H$ behaves according to~\eqref{HinfRW}.\footnote{In practice, to increase the accuracy of the numerical integration, we consider higher-order expansions both near the event horizon and infinity. The coefficients of higher order terms in the asymptotic expansions are given in Appendices~\ref{a:asymptotics} and~\ref{a:horizon}.}
Since it is a homogeneous problem, one of the three free expansion coefficients (one at the horizon, two at infinity) can be fixed without loss of generality.
The first routine is a direct integration method in which we fixed the horizon coefficient to $1$ and numerically integrate the master equation to some large value of the radial coordinate, at which we extract the amplitudes of the normalizable and the non-normalizable solutions. Once those coefficients are determined, we can compute the TLN according to Eq.~\eqref{polarLoveRW}.
The second routine uses instead a shooting method. For definiteness we fix the coefficient of the non-normalizable solution to $C^\text{p}_+ = 1$. We then integrate~\eqref{eq:Polar-AdS} from the horizon to an intermediate radial value $r_\text{int}$ using the horizon expansion, and from infinity (i.e.,\ a sufficiently large value of $r$) to $r_\text{int}$ using the asymptotic expansion. Finally we shoot for the values of the other coefficients such that the solutions determined by the horizon and the asymptotic expansions, and their derivatives, connect smoothly at $r_\text{int}$.
The polar TLN is then computed as in~\eqref{polarLoveRW}.

\begin{figure*}[h!]
\centering
\includegraphics[width=0.45\textwidth]{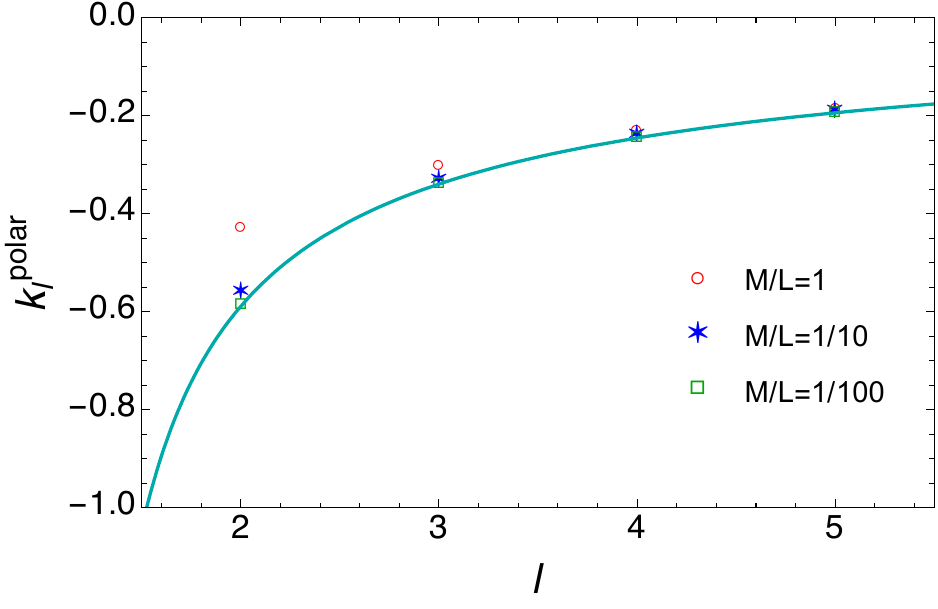}\qquad
\includegraphics[width=0.46\textwidth]{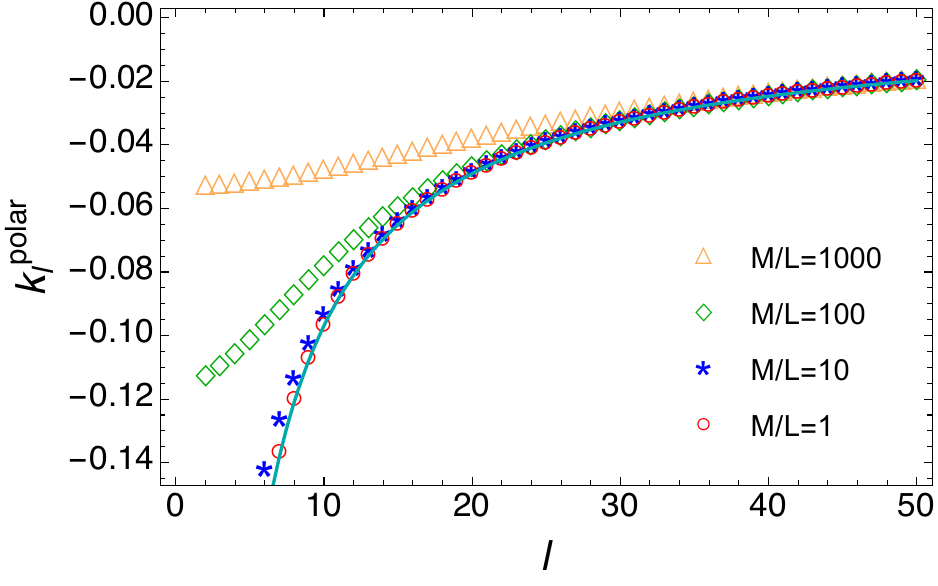}
\caption{Polar TLNs of the Schwarzschild-AdS$_4$ BH computed in the RW formalism for selected values of $M/L\leq1$ (left) and $M/L\geq1$ (right). In both panels, the solid line corresponds to the pure AdS result, see Eq.~\eqref{kRW:pureAdS-polar}, whose large-$l$ behavior is $-1/l$.}
\label{fig:polar}
\end{figure*}

The results obtained are shown in Fig.~\ref{fig:polar} for selected values of $l$ and $M/L$, and reported in Table~\ref{tab} for the lowest harmonic numbers $l$. To better visualize the output, we present small and large masses, in comparison with the AdS length scale, in two distinct panels. Results for larger masses depart more significantly from their pure AdS counterpart, which is denoted by the solid line (see Sec.~\ref{s:PureAdS}), as expected. Independently of the mass, for large $l$ the polar TLNs approach the pure AdS behavior.

\bigskip

%
%%%%%%%%%  FIG 1  %%%%%%%%%
%%%%%%%%%%%%%%%%%%%%%%%%%%%
%

%%%%%%%%%%%%%%%%%%%%%%%%%%%
\subsection{Axial sector}

The asymptotic behavior of the solution to Eq.~\eqref{eq:Axial-AdS} is
\be
h_0 \sim C_+^\text{a} r^2 \left[1 + \frac{3(l-1)(l+2)}{2\Lambda r^2} + \dots \right] + \frac{C_-^\text{a}}{r} \left[ 1 - \frac{3(l-1)(l+2)}{10\Lambda r^2} + \dots \right]\,,
\label{h0infRW}
\ee
and we define the dimensionless axial TLNs as
\be
k_l^\text{axial} \equiv \frac{1}{L^3}\,\frac{C_-^\text{a}}{C_+^\text{a}}\,.
\ee

\begin{figure*}[h!]
\centering
\includegraphics[width=0.435\textwidth]{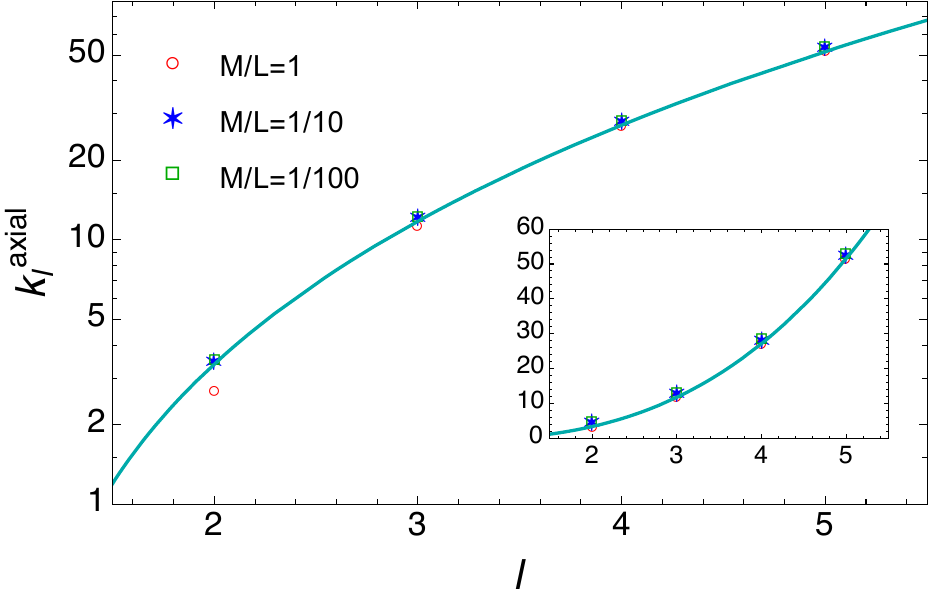}\qquad
\includegraphics[width=0.47\textwidth]{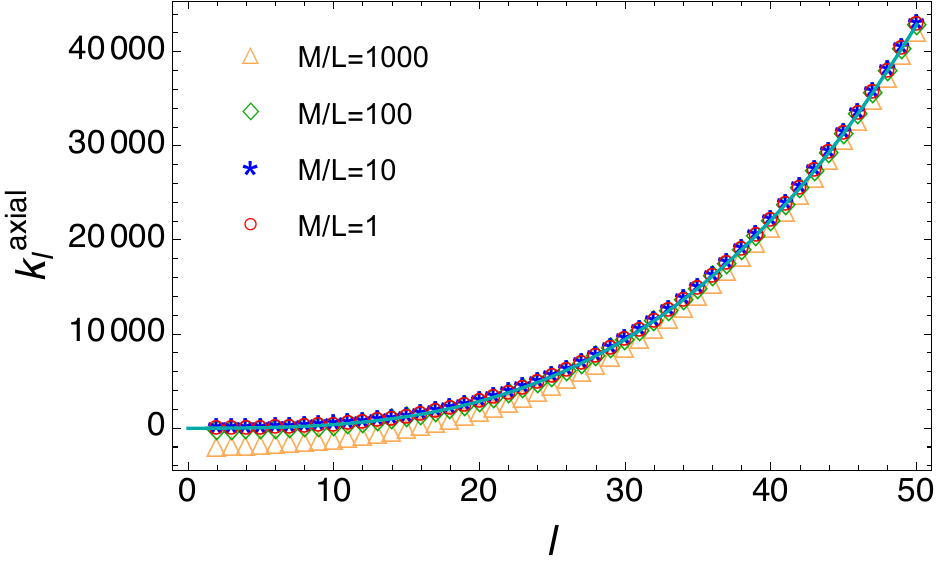}
\caption{Axial TLNs of the Schwarzschild-AdS$_4$ BH computed in the RW formalism for selected values of $M/L\leq1$ (left) and $M/L\geq1$ (right). In both panels, the solid line corresponds to the pure AdS result, see Eq.~\eqref{kRW:pureAdS-axial}, whose large-$l$ behavior is $l^3/3$. Note that for sufficiently large mass, $M$, the low $l$ axial TLNs become negative. A logarithmic scale was used in the left panel to visually enhance departures from the pure AdS result, but the same results in linear scale are displayed in the inset.}
\label{fig:axial}
\end{figure*}

The procedure to compute axial TLNs is analogous to what was described for polar TLNs in the previous subsection~\ref{s:LovePolar}. These results are shown in Fig.~\ref{fig:axial} for selected values of $l$ and $M/L$, and reported in Table~\ref{tab} for the lowest harmonic numbers $l$.

The behavior of the axial TLNs is markedly distinct from the polar sector, even in the pure AdS case (solid line) --- see Sec.~\ref{s:PureAdS}. For sufficiently large mass, the axial TLNs for any multipole $l$ are positive. But, interestingly, for large AdS BHs the TLNs of the lower multipoles become negative in the axial sector.

\bigskip

%
%%%%%%%%%  FIG 2  %%%%%%%%%
%%%%%%%%%%%%%%%%%%%%%%%%%%%
%

%%%%%%%%%  TABLE 1  %%%%%%%%%
\begin{table}[t!]
\centering%
\begin{tabular}{cdddddd}
%used to be {ccccccc}
\toprule
$M/L$ & \multicolumn{2}{c}{0.1} & \multicolumn{2}{c}{1} & \multicolumn{2}{c}{10} \\\midrule
& \mc{polar} & \mc{axial} & \mc{polar} & \mc{axial} & \mc{polar} & \mc{axial} \\\midrule
$l=2$ & -0.5639 &   3.3462 & 
        -0.4320 &   2.6296 & 
        -0.2347 & -11.4815 \\
$l=3$ & -0.3355 &  11.7197 & 
        -0.3060 &  11.0703 & 
        -0.2089 &  -0.8597 \\
$l=4$ & -0.2442 &  27.0959 & 
        -0.2342 &  26.4547 & 
        -0.1839 &  16.2386 \\
$l=5$ & -0.1935 &  51.4725 & 
        -0.1892 &  50.8272 & 
        -0.1617 &  41.8260 \\
    \bottomrule
\end{tabular}
\caption{Values of the TLNs of the Schwarzschild-AdS$_4$ BH computed in the RW formalism for selected values of $M/L$ and for the lowest multipoles. These numbers were obtained with the accuracy shown using both the direct integration and with the shooting routines. For polar TLNs it was typically possible to achieve accuracy down to the seventh decimal place.
}\label{tab}
\end{table}
%%%%%%%%%%%%%%%%%%%%%%%%%%%

%%%%%%%%%%%%%%%%%%%%%%%%%%%
\subsection{Pure AdS and eikonal limit\label{s:PureAdS}}

For pure AdS ($M=0$) the TLNs can be computed analytically.
In the RW language they are 
\be
\label{kRW:pureAdS-polar}
k_l^\text{polar} &= -\frac{l(l+1)}{2(l-1)(l+2)}\frac{\Gamma \left(\frac{l+1}{2}\right)^2}{\Gamma \left(\frac{l}{2}+1\right)^2}
\; \stackrel{l\to\infty}{\longrightarrow} \; - \frac{1}{l}\,, \\
\label{kRW:pureAdS-axial}
k_l^\text{axial} &= \frac{2(l-1)(l+2)}{3}\frac{\Gamma \left(\frac{l}{2}+1\right)^2}{\Gamma \left(\frac{l+1}{2}\right)^2}
\; \stackrel{l\to\infty}{\longrightarrow} \; \frac{l^3}{3}\,,
\ee
where we have taken the eikonal limit in the last step.
These results are plotted in Figs.~\ref{fig:polar} and~\ref{fig:axial} as solid lines to help guide the eye, even though $l$ should be understood as a discrete quantity.

To better understand why all the TLNs tend to the pure AdS result when $l\to\infty$, it is useful to write the master equations in terms of the dimensionless variable $\rho$ and parameter $\mu$. For the axial sector, for example,
\be
\rho^2 h_0''(\rho) - \frac{2\rho^2+l(l+1)-\frac{4\mu}{\rho}}{\rho^2+1-\frac{2\mu}{\rho}} h_0(\rho) = 0\,.
\ee
Introducing a new rescaled variable $y\equiv (\rho-\rho_h)/l^2$, which takes values in the interval $[0,\infty)$ \textit{independently} of $\rho_h(\mu)$ for the region of interest to us ---the exterior of the BH---, one finds
\be
\left[y^2 + O(l^{-2})\right] h_0''(y) - \left[2+\frac{l(l+1)}{l^4 y^2} + O(l^{-4}) \right] h_0(y) = 0\,,
\ee
where, $\rho_h(\mu)$ is a function of the mass parameter and it indicates the horizon location in terms of the $\rho$ coordinate.

Therefore, up to order $O(l^{-2})$, the master equation does not depend on the mass parameter at all. The horizon radius $\rho_h$ makes its first appearances in the subleading terms not shown explicitly.  The upshot is that for large $l$ the equation to solve, together with the boundary conditions (which in these coordinates are unchanged), does not depend on the mass and must therefore agree with the result for $M=0$, i.e., pure AdS. A similar story applies to the polar sector.

At the intuitive, more physical, level, large $l$ means modes with large wavenumber. In AdS, such modes live close to its conformal boundary and are therefore insensitive to the infrared, deep bulk, dynamics. So, it should not make a difference for the large $l$ TLNs if they are being computed in pure AdS or with a BH in its interior.

%%%%%%%%%%%%%%%%%%%%%%%%%%%
\subsection{Schwarzschild limit}

When the cosmological constant identically vanishes, the spacetime changes its asymptotic structure and reduces to the Schwarzschild BH.

The equations governing static gravitational perturbations can be readily obtained by taking $\Lambda=0$ in Eqs.~\eqref{eq:Polar-AdS} and~\eqref{eq:Axial-AdS}. Nonetheless, for a Schwarzschild background the exponents of the leading terms in the asymptotic solutions depend on $l$, namely as $r^{l+1}$ and $1/r^l$ in both sectors. Hence, the asymptotically flat solutions are \emph{not} the $\Lambda\to0$ limit of those obtained for a Schwarzschild-AdS BH.
As a consequence, also the relations among TLNs will be different.

Indeed, for a Schwarzschild background, the perturbative equations can be solved exactly, and by imposing regularity of the solutions on the event horizon one finds that TLNs are identically zero, in both sectors, for any value of $l$~\cite{Binnington:2009bb,Damour:2009vw},
\be
k_l^\text{polar} = k_l^\text{axial} = 0\,.
\ee

%%%%%%%%%%%%%%%%%%%%%%%%%%%
%%%%%%%  SECTION 4  %%%%%%%
%%%%%%%%%%%%%%%%%%%%%%%%%%%
\section{Love numbers of Schwarzschild-AdS\texorpdfstring{$_4$}{4} in the Kodama-Ishibashi formalism\label{s:LoveKI}}
%%%%%%%%%%%%%%%%%%%%%%%%%%%

In Sec.~\ref{s:LoveSAdS4}, the equations were obtained following a gauge-fixed approach, namely by adopting the so-called RW gauge~(\ref{heven}--\ref{hodd}). However, it is also possible to work explicitly with gauge-invariant quantities by employing the formalism developed in~\cite{Kodama:2003jz,Ishibashi:2011ws}.
While in practice there is no formal difference, using a gauge-invariant definition of the TLNs (namely, the KI formalism) allows us to confirm the numerical results of the previous section and, in addition, to derive relations between TLNs in the different formalisms. 

In this section we begin by presenting the relevant equations that govern gravitational perturbations of Schwarzschild-AdS$_4$ in the KI gauge-invariant approach, and relate them with our RW gauge-fixed expressions.

In the KI formalism, the master perturbation equation for the scalar (or polar) sector, expressed in terms of the gauge-invariant master variable $\Phi_S$, reads (see~\cite{Kodama:2003jz} for details)
\be
\label{eq:MasterScalar}
f(r) \frac{d}{dr}\left[ f(r) \frac{d\Phi_S}{dr} \right] 
- \frac{f(r)}{r^2} \left[\frac{(\mathfrak{m}+2)\mathfrak{m}^2 + 3\mathfrak{m}^2 x + 9(\mathfrak{m}-2y)x^2 + 9x^3}{\left(\mathfrak{m}+3x\right)^2} \right]\Phi_S = 0\,,
\ee
where $f(r)=1 - 2M/r - \Lambda r^2/3 = e^{-\ol{\lambda}(r)}$ is the blackening factor~\eqref{eq:bkgd_metric} and, in accordance with~\cite{Kodama:2003jz}, we defined $\mathfrak{m}\equiv (l-1)(l+2)$, $x\equiv 2M/r$ and $y\equiv \Lambda r^2/3$ for convenience. The quantity $\mathfrak{m}$ should not be confused with the azimuthal number $m$.

It can be checked that~\eqref{eq:MasterScalar} is equivalent to (a derivative of) Eq.~\eqref{eq:Polar-AdS}. 
Concretely the mapping that relates the two equations is
\be
\Phi_S &= 2r \left\{
\frac{\mathfrak{m}^2 (x + y - 1)+\mathfrak{m} [9xy + x - 2 (y+1)] + 3x [x (8y - 1) - 2y (y+1)]}{f(r)\mathfrak{m} (\mathfrak{m}+2)(\mathfrak{m}+3x)}
\right\}H \nn\\
&+ 2r^2 \left\{ \frac{ 2\mathfrak{m} - 3\mathfrak{m} x + 6yx - 3x^2}{  (\mathfrak{m}+2) \mathfrak{m} \left(\mathfrak{m}+3x\right)} \right\} H' \,.
\label{eq:MappingScalar}
\ee
These expressions are valid for all integers $l\geq2$.

Regarding the vector (or axial) sector, the KI master perturbation equation is
\be
f(r) \frac{d}{dr}\left[ f(r) \frac{d\Phi_V}{dr} \right] - \frac{f(r)}{r^2} \left[ l(l+1) - \frac{6M}{r} \right] \Phi_V = 0\,.
\label{eq:MasterVector}
\ee
Similarly to the polar sector,~\eqref{eq:Axial-AdS} is also equivalent to a derivative of~\eqref{eq:MasterVector} under the following transformation:
\be
h_0 = -f(r) \frac{d}{dr}\left( r\, \Phi_V \right).
\label{eq:MappingVector}
\ee

The transformations~\eqref{eq:MappingScalar} and~\eqref{eq:MappingVector} can be used to relate the master equations in the RW and in the KI formalisms. These relations are somewhat lengthy and not particularly enlightening, so we relegate them to Appendix~\ref{a:master_relations}.

With these relations at hand, once we know the asymptotic solutions of the master perturbation equations, we are able to compare the TLNs computed in the RW formalism to those computed in the KI formalism.

%%%%%%%%%%%%%%%%%%%%%%%%%%%
\subsection{Scalar (polar) sector}

The asymptotic behavior of the solution to Eq.~\eqref{eq:MasterScalar}, including explicitly terms up to order $O(r^{-3})$, takes the form
\be
\Phi_S &\sim  D_+^\text{s} \left[ 1+\left(\frac{36M^2}{\mathfrak{m}^2}-\frac{3(\mathfrak{m}+2)}{2\Lambda}\right)\frac{1}{r^2}+\dots \right] + D_-^\text{s} \left[ \frac{1}{r}  +\dots \right]\,.
\label{eq:asymptoticsPhiS}
\ee
Using the mapping~\eqref{eq:MappingScalar} we can relate the two sets of expansion coefficients:
\be
C_-^\text{p} &= \left[ \frac{12 \Lambda  M^2}{\mathfrak{m}^2} -\frac{\mathfrak{m}+2}{2}\right] D_+^\text{s} + \frac{2  \Lambda  M}{\mathfrak{m}} D_-^\text{s}\,,\\
C_+^\text{p} &= -3M D_+^\text{s} -  \frac{\mathfrak{m}}{2} D_-^\text{s}  \,.
\ee
Note that the non-normalizable coefficient $C_-^\text{p}$ and the normalizable coefficient $C_+^\text{p}$ get mixed as they are mapped into $D_+^\text{s}$ and $D_-^\text{s}$.

Expressed in terms of the asymptotic behavior of $\Phi_S$, the polar TLNs in the RW formalism~\eqref{polarLoveRW} become
\be\label{polarRWtoKI}
k^\text{polar}_l = \frac{ l(l+1) + \frac{12M}{(l-1)(l+2)L} \left( \mathcal{K}^S_l +\frac{6M}{(l-1)(l+2)L} \right)}{(l-1)(l+2) \mathcal{K}^S_l + \frac{6M}{L}}\,,
\ee
where we have defined the KI (gauge-invariant) TLNs for the scalar sector as
\be
\mathcal{K}^S_l \equiv \frac{1}{L}\frac{D_-^\text{s}}{D_+^\text{s}} \,.
\ee
Eq.~\eqref{polarRWtoKI} can be more compactly written as
\be\label{polarRWtoKIcomp}
k^\text{polar}_l = \frac{ \mathfrak{m}+2 + \frac{12\mu}{\mathfrak{m}} \left( \mathcal{K}^S_l +\frac{6\mu}{\mathfrak{m}} \right) }{\mathfrak{m} \left( \mathcal{K}^S_l + \frac{6\mu}{\mathfrak{m}} \right) }\,.
\ee

Observe that the limit $M\to0$ yields 
\be
k^\text{polar}_l \to \frac{l(l+1)}{(l-1)(l+2)\,\mathcal{K}^S_l}.
\ee
Note that naively the limit $\Lambda\to0$ would give the same result but, as we comment below, this asymptotically flat limit should not be taken directly in this result because the master equations, and their solutions, with $\Lambda\neq0$ and with $\Lambda=0$ have different natures. So, the limit $\Lambda\to0$ of the TLNs cannot be taken continuously.
Moreover, in the eikonal limit $l\to\infty$ we have $k^\text{polar}_l \to (\mathcal{K}^S_l)^{-1}$, irrespective of the values taken by $M, \Lambda$.

\begin{figure*}[t!]
\centering
\includegraphics[width=0.45\textwidth]{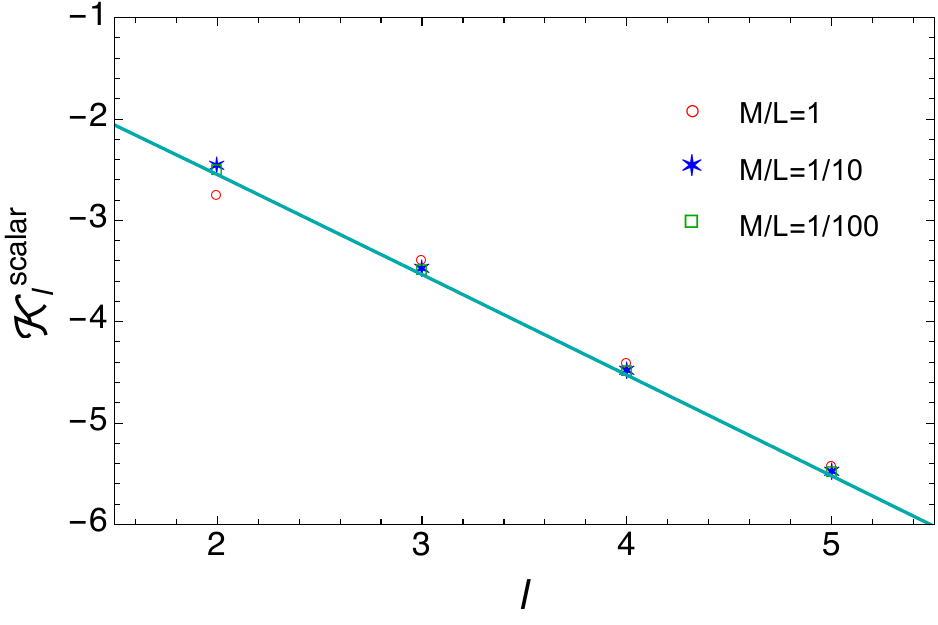}\qquad
\includegraphics[width=0.46\textwidth]{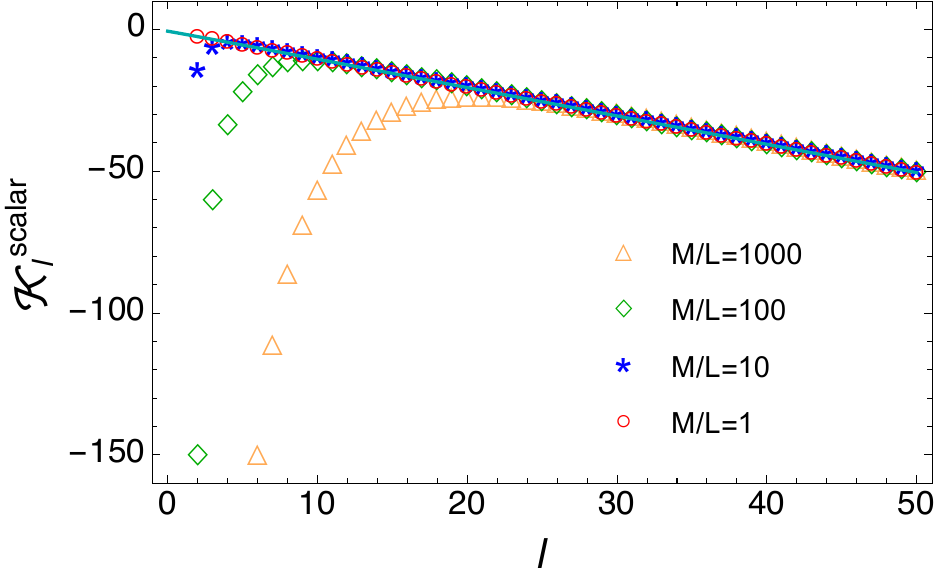}
\caption{Scalar TLNs of the Schwarzschild-AdS$_4$ BH computed in the KI formalism for selected values of $M/L\leq1$ (left) and $M/L\geq1$ (right). In both panels, the solid line corresponds to the pure AdS result, see \eqref{kKI:pureAdS}, whose
large-l behavior is $-l$.}
\label{fig:scalar_KI}
\end{figure*}

Working in the KI formalism, one can compute the TLNs $\mathcal{K}^S_l$ by numerically integrating~\eqref{eq:MasterScalar} in a manner similar to what was described in~\ref{s:LovePolar}, for various values of the multipolar number $l$ and the ratio $M/L$. 
The results obtained are shown in Fig.~\ref{fig:scalar_KI} and in Table~\ref{tab2} for selected values of $M/L$ and $l$. The values for the TLNs $\mathcal{K}^S_l$ are in excellent agreement with those shown in Fig.~\ref{fig:polar} for the quantity $k_l^\text{polar}$, once the conversion formula~\eqref{polarRWtoKI} is used.

It is worth remarking that, as long as $\mu=M/L\neq0$, Eq.~\eqref{polarRWtoKIcomp} implies that a vanishing (scalar) KI TLN corresponds to a non-zero (polar) RW TLN, and vice-versa.

%%%%%%%%%%%%%%%%%%%%%%%%%%%%%%%%%%%%%
%%%%%%%%%  FIG 3 Scalar KI  %%%%%%%%%
%%%%%%%%%%%%%%%%%%%%%%%%%%%

%%%%%%%%%%%%%%%%%%%%%%%%%%%%%%%%%%%%%
%%%%%%%%%  FIG 4 Vector KI  %%%%%%%%%
\begin{figure*}[t!]
\centering
\includegraphics[width=0.45\textwidth]{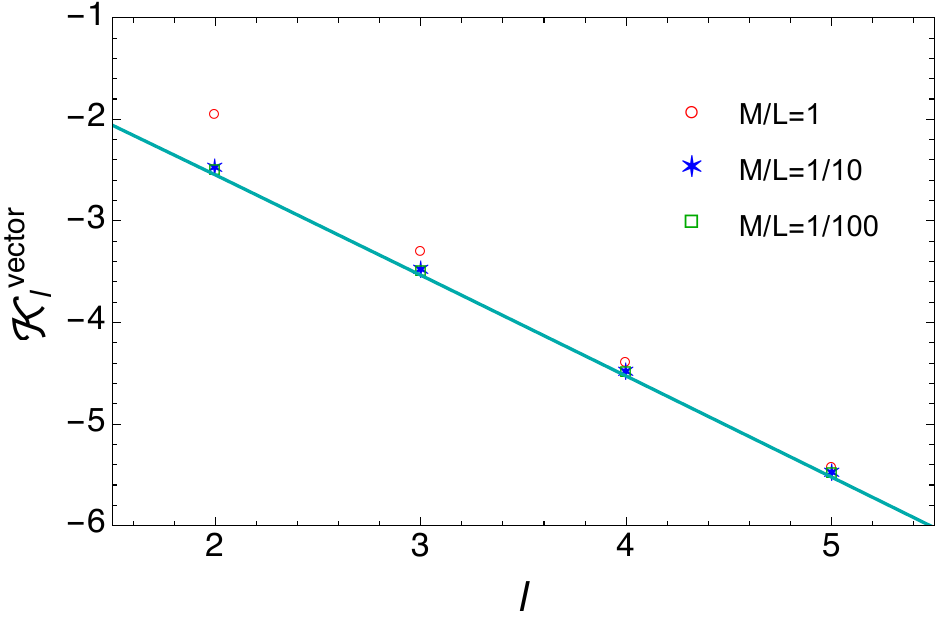}\qquad
\includegraphics[width=0.46\textwidth]{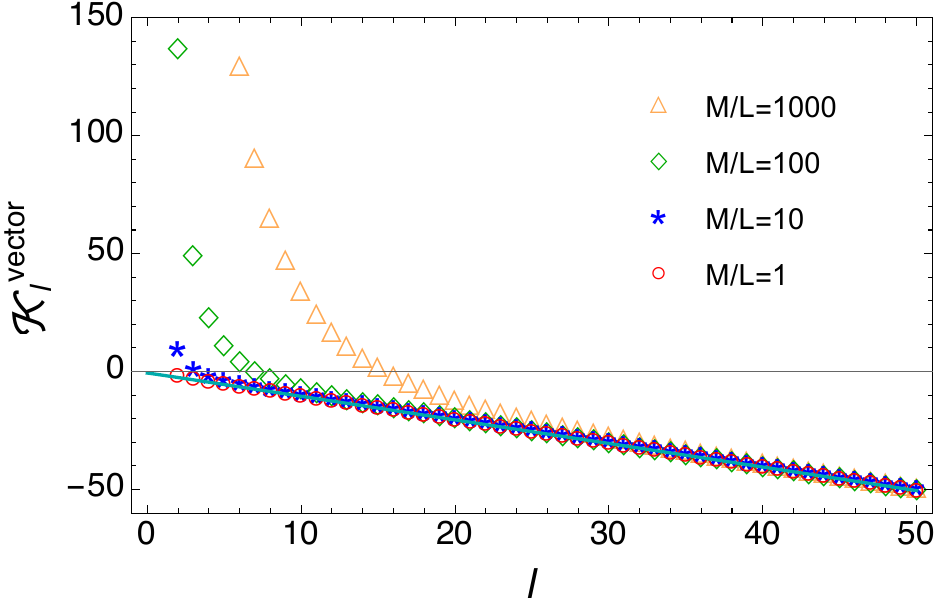}
\caption{Vector TLNs of the Schwarzschild-AdS$_4$ BH computed in the KI formalism for selected values of $M/L\leq1$ (left) and $M/L\geq1$ (right). In both panels, the solid line corresponds to the pure AdS result, see \eqref{kKI:pureAdS}, whose
large-l behavior is $-l$.}
\label{fig:vector_KI}
\end{figure*}
%%%%%%%%%%%%%%%%%%%%%%%%%%%

%%%%%%%%%%%%%%%%%%%%%%%%%%%
\subsection{Vector (axial) sector}

The asymptotic behavior of the solution to Eq.~\eqref{eq:MasterVector} is
\be\label{eq:asy_expansion_vector}
\Phi_V &\sim D_+^\text{v} \left[ 1-\frac{3(\mathfrak{m}+2)}{2\Lambda r^2} +\dots \right] + \frac{D_-^\text{v}}{r} \left[ 1 - \frac{\mathfrak{m}}{2\Lambda r^2} + \frac{3\mathfrak{m}(\mathfrak{m}-10)}{40\Lambda^2 r^4} +\dots \right].
\ee

The mapping~\eqref{eq:MappingVector} allows one to relate the two sets of expansion coefficients as follows,\footnote{Under the transformation~\eqref{eq:MappingVector}, the subleading terms shown in expansion~\eqref{eq:asy_expansion_vector} get appropriately mapped to the subleading terms displayed in expansion~\eqref{h0infRW}.}
\be
C_+^\text{a} &= \frac{\Lambda}{3} D_+^\text{v} \,, \\
C_-^\text{a} &= \frac{(l-1)(l+2)}{3} D_-^\text{v} \,,
\ee
so that, expressed in terms of the asymptotic behavior of $\Phi_V$, the axial TLNs become
\be\label{axialRWtoKI}
k^\text{axial}_l = -\frac{(l-1)(l+2)}{3} \mathcal{K}^V_l\,,
\ee
where
\be
\mathcal{K}^V_l \equiv \frac{1}{L}\frac{D_-^\text{v}}{D_+^\text{v}}
\ee
denote the KI TLNs for the vector sector.
In this sector the TLNs in both formalisms are proportional to each other, so the vanishing of one of them implies the vanishing of the other one as well. 

Again, the results we obtained by numerically integrating~\eqref{eq:MasterVector} for a range of values of the multipole number $l$ and the quantity $M/L$ ---see Fig.~\ref{fig:vector_KI} and Table~\ref{tab2}--- are in precise agreement with those shown in Fig.~\ref{fig:axial} and Table~\ref{tab}, upon using the conversion formula~\eqref{axialRWtoKI}.

%%%%%%%%%  TABLE 2  %%%%%%%%%
\begin{table}[t!]
\centering%
\begin{tabular}{cdddddd}
\toprule
$M/L$ & \multicolumn{2}{c}{0.1} & \multicolumn{2}{c}{1} & \multicolumn{2}{c}{10} \\\midrule
& \mc{scalar} & \mc{vector} & \mc{scalar} & \mc{vector} & \mc{scalar} & \mc{vector} \\\midrule
$l=2$ & -2.4975 & -2.5097 & 
        -2.7690 & -1.9722 & 
       -15.1939 &  8.6111 \\
$l=3$ & -3.5124 & -3.5159 & 
        -3.4166 & -3.3210 & 
        -6.8516 &  0.2579 \\
$l=4$ & -4.5147 & -4.5159 & 
        -4.4284 & -4.4091 & 
        -5.3377 & -2.7064 \\
$l=5$ & -5.5143 & -5.5149 & 
        -5.4509 & -5.4457 & 
        -5.5463 & -4.4813  \\\bottomrule
\end{tabular}
\caption{Values of the TLNs of the Schwarzschild-AdS$_4$ BH computed in the KI formalism for selected values of $M/L$ and for the lowest multipoles.
These numbers were obtained with (at least) the accuracy shown using both the direct integration and with the shooting routines. In most cases it was possible to achieve accuracy down to the ninth decimal place.  
These values are also in excellent agreement with those shown in Table~\ref{tab}:\ once the conversion formulas~\eqref{polarRWtoKI} and~\eqref{axialRWtoKI} are used to map the results between the different formalisms, the relative errors obtained are all smaller than 0.5\%.
}
\label{tab2}
\end{table}
%%%%%%%%%%%%%%%%%%%%%%%%%%%

%%%%%%%%%%%%%%%%%%%%%%%%%%%
\subsection{Pure AdS and eikonal limit}

For pure AdS, in the KI formalism the TLNs have the same value in both sectors
\be\label{kKI:pureAdS}
\mathcal{K}_l^{S/V} = -\frac{2\,\Gamma \left(\frac{l}{2}+1\right)^2}{\Gamma \left(\frac{l+1}{2}\right)^2}
\; \stackrel{l\to\infty}{\longrightarrow} \; - l \,.
\ee
It is easy to verify that \eqref{kKI:pureAdS} and (\ref{kRW:pureAdS-polar}--\ref{kRW:pureAdS-axial}) are mapped into each other through \eqref{polarRWtoKI} and \eqref{axialRWtoKI} with $M=0$.

Note that this is an interesting property specific to the KI formalism and it applies only to the pure AdS background -- finite mass breaks this symmetry but it is an approximate symmetry for  small BHs in AdS. We can see from Figs.~\ref{fig:scalar_KI} and~\ref{fig:vector_KI} that this is the case for very small values of the BH masses.
This property is reminiscent of the isospectrality of quasi-normal modes of Schwarzschild BHs in GR in RW formalism~\cite{Chandrasekhar:1975zza,Glampedakis:2017rar}.

%%%%%%%%%%%%%%%%%%%%%%%%%%%
\subsection{Schwarzschild limit}

Even in the KI formalism, when we take $\Lambda=0$ in the scalar and vector master equations~\eqref{eq:MasterScalar} and \eqref{eq:MasterVector} we obtain the equations for a Schwarzschild background.
But also in this case, the nature of the solutions is different, and we cannot neither take the $\Lambda\to0$ limit of the asymptotic expansions~\eqref{eq:asymptoticsPhiS} and \eqref{eq:asy_expansion_vector}, nor of the relations~\eqref{polarRWtoKI} and \eqref{axialRWtoKI} among TLNs in the RW and KI formalisms.

Remarkably, when $\Lambda=0$ the master scalar and vector equations can be solved exactly in a closed form.
Full details are given in Appendix~\ref{a:LoveKISchwarz}, but the result is that even in the KI formalism, TLNs of a Schwarzschild BH are identically zero, in both sectors, for any value of $l$,
\be
\mathcal{K}_l^S = \mathcal{K}_l^V = 0\,.
\ee

%%%%%%%%%%%%%%%%%%%%%%%%%%%
%%%%%%%  SECTION 5  %%%%%%%
%%%%%%%%%%%%%%%%%%%%%%%%%%%
\section{The holographic dual of Love numbers:\ geometric polarization\label{s:holoLove}}  
%%%%%%%%%%%%%%%%%%%%%%%%%%%

The AdS/CFT correspondence, also known as gravity-gauge holography, is an important and revolutionary connection between gravitational theories and quantum field theories that has been heavily explored in the last decades.
In this framework a (large) Schwarzschild-AdS$_4$ BH with temperature~$T$ is holographically dual to a conformal plasma living on the boundary of AdS, at the same temperature~\cite{Witten:1998qj, Marolf:2013ioa}. The conformal boundary of the unperturbed bulk is simply $\mathbb{R}\times S^2$, but once gravitational perturbations are included, the spatial 2-sphere gets correspondingly deformed.

Ref.~\cite{Emparan:2017qxd} put forward the holographic dual interpretation of TLNs of AdS BHs:\ in the dual CFT they are regarded as coefficients that control  geometric polarization effects.
One can be more explicit, by computing the stress-energy tensor for the dual CFT, as well as the induced metric on the boundary, determined by the Schwarzschild-AdS perturbed by a chosen tidal field. The term ``geometric polarization''  is meant to express the non-trivial way in which the components of the CFT stress-energy tensor adapt to changes of the geometry of the space on which the CFT lives. In our context, the TLNs computed dictate how the stress-tensor responds to small deformations of the 2-sphere.

At this point, it is useful to fix some notation. So far we have adopted the greek letters  $\alpha, \beta, \mu,\nu$ as indices for the AdS bulk coordinates, e.g., $x^\mu = \{t,r,\theta,\phi\}$. We shall instead use letters from the beginning of the latin alphabet, $a,b,\dots$, for indices of the coordinates parametrizing the dual field theory directions, e.g., $x^a = \{t,\theta,\phi\}$.
To be concrete, it is convenient to perform a change of the radial coordinate $r$ to use instead the coordinate $v$ parametrizing the holographic direction on AdS, and in such a way that the boundary of AdS lies at $v=0$. They are simply related through
\be
r=\frac{L^2}{v}\,.
\ee
This choice of radial coordinate is by no means mandatory (see~\cite{Bhattacharyya:2007vs}, for example), but it allows to connect more cleanly with the analysis of~\cite{Emparan:2017qxd}.
Expressed in terms of $v$, the unperturbed Schwarzschild-AdS line element is written as
\be
ds^2 = \frac{L^2}{v^2}\left[- F(v) dt^2 + \frac{dv^2}{F(v)} + L^2 d\Omega_2^2\right]\,,
\label{eq:line_elem_unpert}
\ee
where we defined
\be
F(v) = 1 + \frac{v^2}{L^2} - \frac{2Mv^3}{L^4}\,.
\ee
Naturally, perturbations (either polar or axial) can be equally expressed in terms of $v$. Note, however, that even though the bulk metric is static and spherically symmetric --- and in particular displays no explicit dependence on coordinates $t$ and $\phi$ ---, the (static) perturbations we consider generically break spherical symmetry. So our perturbed bulk metric depends on coordinates $\{v,\theta,\phi\}$.

Specifically, the bulk line element perturbed with a polar mode gets the following contributions, dictated by~\eqref{heven}, in addition to~\eqref{eq:line_elem_unpert}:
\be
+ \frac{L^2}{v^2} \left[- \frac{H(v)}{F(v)} dv^2 - H(v) F(v) dt^2 + L^2 K(v) d\Omega_2^2 \right] Y^{lm}(\theta,\phi)\,,
\ee
with the understanding that the function $K$ is expressed in terms of $H$ and its derivative through constraint~\eqref{eq:Kconstraint}.
For a mode in the axial sector the additional terms, from~\eqref{hodd}, read
\be
- 2h_0(v) \left(S_\theta^{lm} d\theta - S_\phi^{lm} d\phi \right) dt\,.
\ee

One then proceeds by computing the metric induced from the bulk (including perturbations from the tidal field and from the induced multipole moments) on a (2+1)-dimensional constant-$v$ slice. This induced metric is denoted by $\widehat{g}_{ab}(v,\theta,\phi)$. The renormalized boundary metric $\gamma_{ab}$ is then obtained by
\be
\gamma_{ab}(\theta,\phi) = \lim_{v\to0} \left(\frac{v^2}{L^2}\right)\widehat{g}_{ab}(v,\theta,\phi)\,.
\ee

In the polar sector we find that the boundary metric is diagonal,
\be
\gamma_{ab}dx^a dx^b =-dt^2 + \left(L^2+\frac{2 C^{\rm p}_+ Y^{lm}(\theta,\phi)}{(l+2)(l-1)}\right)d\Omega_2^2\,.
\ee
The fact that $C^{\rm p}_+$, the coefficient of the subdominant term in the asymptotic expansion of $H$ in \eqref{HinfRW}, appears in the boundary metric, instead of $C^{\rm p}_-$, can be confirmed by mapping the non-normalizable coefficient in the KS scalar channel master variable first into the KI $D^{\rm s}_+$ and $D^{\rm s}_-$ coefficients, and from those to the RW $C^{\rm p}_+$ and $C^{\rm p}_-$. See Appendix~\ref{s:LoveKS} for further details about this conversion.

In the axial sector the boundary metric reads
\be
\gamma_{ab}dx^a dx^b = - dt^2 + L^2 d\Omega_2^2
+ 2L^2 C^{\rm a}_+ \left( S_\phi^{lm}\,d\phi + S_\theta^{lm}\,d\theta \right)dt\,,
\ee
and performing the same mapping of the non-normalizable coefficients between the different formalisms as we did in the polar sector, we find that KS vector channel master variable is directly proportional to $C^{\rm a}_+$, as expected.

The non-normalizable behavior of the perturbed bulk metric dictates the boundary metric and therefore also the curvature of the spacetime in which the CFT lives. On the other hand, from the normalizable solutions of the bulk metric one reads off the corrections to the CFT stress-energy tensor. The TLNs thus contain information about how the properties (energy density, momentum, pressure, shear) of the plasma on the boundary change if one deforms the geometry in which it lives.

In the case we have focused on, the CFT lives on a 2-sphere (plus time, but it plays no role herein because we are considering static TLNs). We consider that the sphere departs slightly from spherical symmetry, and then calculate how much the plasma properties change accordingly, using gravitational linear perturbations in the bulk and then translating the results to the boundary stress-energy tensor according to the AdS/CFT dictionary~\cite{Balasubramanian:1999re, deHaro:2000vlm},
\be
T_{ab} &= \lim_{v\to0} \frac{L}{v} \widehat{T}_{ab}\,,\\
8\pi {\cal G}_3\, \widehat{T}_{ab} &= \widehat{K}_{ab} - \widehat{K}\widehat{g}_{ab} - \frac{2}{L}\widehat{g}_{ab} + L\widehat{G}_{ab}\,.
\ee
Here, $\widehat{K}_{ab}$ denotes the extrinsic curvature of a $v=\text{constant}$ bulk hypersurface, $\widehat{K}$ is its trace, and $\widehat{G}_{ab}$ is the Einstein tensor of the induced metric $\widehat{g}_{ab}$ on the same hypersurface. ${\cal G}_3$ refers to the three-dimensional Newton constant.

Following this procedure for the unperturbed AdS-Schwarzschild bulk metric one finds a non-trivial boundary stress-energy tensor corresponding to a uniform plasma on the sphere at temperature $T\propto M^{1/3}$. Linear gravitational perturbations in each sector induce non-uniform corrections responsible for the geometric polarization,
\be \label{eq:stress_tot}
T_{ab} = T_{ab}^{(0)} + \sum_{l,m}\delta T_{ab}^{lm,\rm axial} + \sum_{l,m}\delta T_{ab}^{lm,\rm polar}\,,
\ee
where $T_{ab}^{(0)}$ represents the unperturbed stress-energy tensor, 
\be \label{eq:stress_zero}
T_{ab}^{(0)} =
\frac{M}{8\pi{\cal G}_3 L^2} \text{diag} \{2,L^2,L^2\sin^2\theta\}\,.
\ee
The contribution of the polar sector perturbations to the resulting boundary stress-energy tensor is 
\be \label{eq:stress_polar}
\delta T_{ab}^{lm,\rm polar} =
\frac{M}{8\pi{\cal G}_3 L^2}
\left(
\begin{array}{ccc}
 \xi_1 Y^{lm} & 0 & 0 \\
 0 & L^2\xi_2 & L^2\xi_4 \\
 0 & L^2\xi_4 & L^2\xi_3 \\
\end{array}
\right)\,,
\ee
whereas in the axial sector the corrections appear only in off-diagonal terms,
\be
\delta T_{ab}^{lm,\rm axial} =
\frac{M}{8\pi{\cal G}_3 L^2}
\left(
\begin{array}{ccc}
 0 & \chi L S_\theta^{lm} & \chi L S_\phi^{lm} \\
 \chi L S_\theta^{lm} & 0 & 0 \\
 \chi L S_\phi^{lm}   & 0 & 0 \\
\end{array}
\right)\,.
\ee
We have defined the following dimensionless combinations for convenience:
\be
\xi_1 &= -\frac{C^{\rm p}_+}{L^2} \frac{48 + \frac{L k_l^{\rm polar}}{2M} \left(3 \mathfrak{m}^2 - 20\mathfrak{m}+48\right)}{\mathfrak{m}}\,,\label{eq:xi_1}\\
\xi_2(\theta,\phi) &= \frac{C^{\rm p}_+}{L^2}
\left[\frac{26 + \frac{L k_l^{\rm polar}}{2M} \left(\mathfrak{m}^2-9 \mathfrak{m}+24\right)}{\mathfrak{m}}\,Y^{lm}
+\frac{L k_l^{\rm polar}}{2M} \csc^2\theta \left(\cos\theta\, S^{lm}_\phi - m^2\, Y^{lm}\right) \right],\\
\xi_3(\theta,\phi) &= \frac{C^{\rm p}_+}{L^2}
\left[\frac{26 - \frac{L k_l^{\rm polar}}{2M} \left(11 \mathfrak{m}-24\right)}{\mathfrak{m}}\,\sin^2\theta\, Y^{lm}
-\frac{L k_l^{\rm polar}}{2M} \left(\cos\theta\, S^{lm}_\phi -  m^2\, Y^{lm}\right)\right],\\
\xi_4(\theta,\phi) &= -\frac{C^{\rm p}_+}{L^2} \frac{L k_l^{\rm polar}}{2M}
\left( Y^{lm}_{,\theta,\phi} + S^{lm}_\theta \cos\theta \right),\\
\chi &= C^{\rm a}_+ L \left[1 + \frac{3 L k_l^{\rm axial}}{2M}\right].
\ee

%%%%%%%%%  FIG 5  %%%%%%%%%
\begin{figure*}[t!]
\centering
\!\!\includegraphics[width=0.45\textwidth]{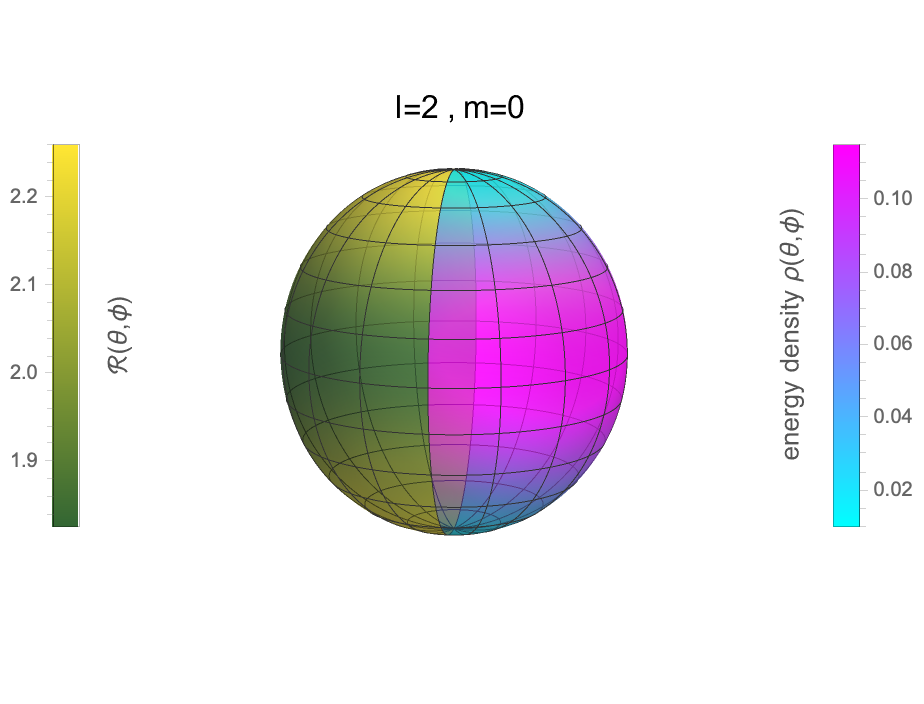}\qquad\;
\includegraphics[width=0.46\textwidth]{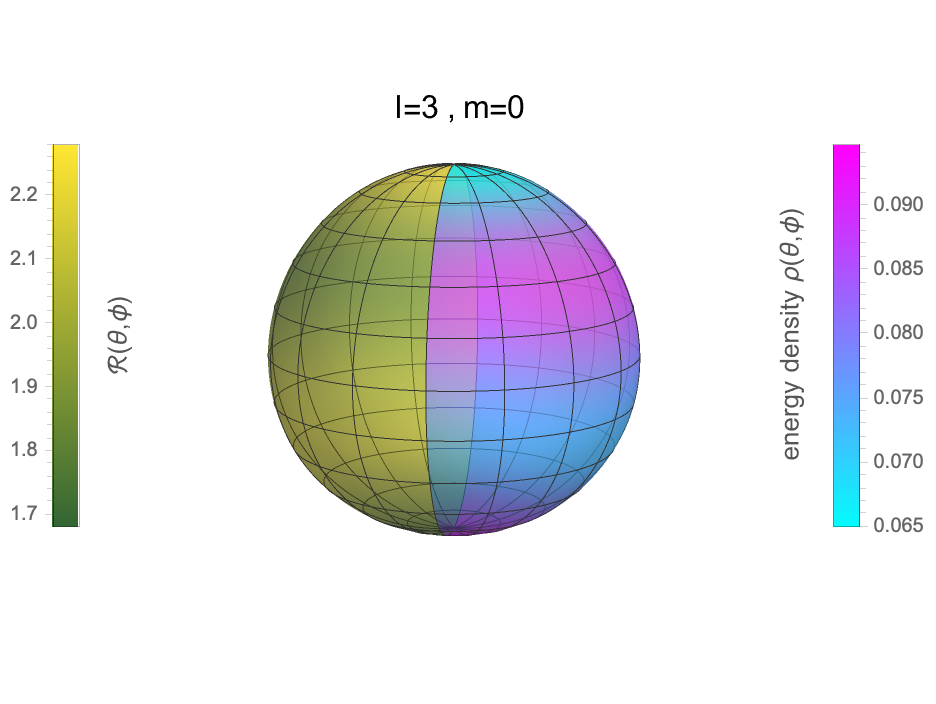}\\
\vspace{-1.0cm}
\includegraphics[width=0.45\textwidth]{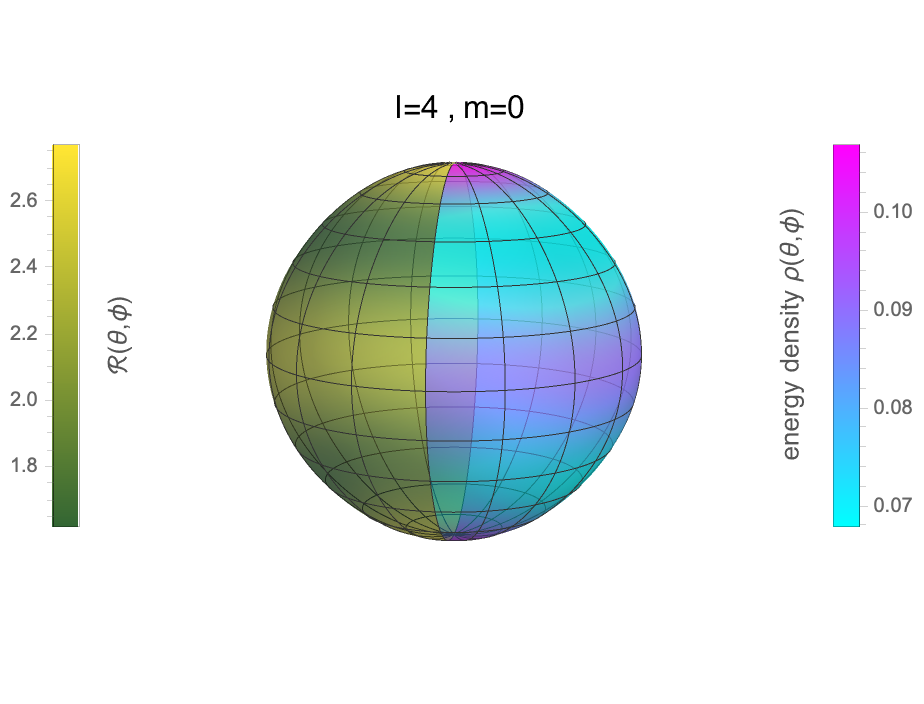}\qquad
\includegraphics[width=0.46\textwidth]{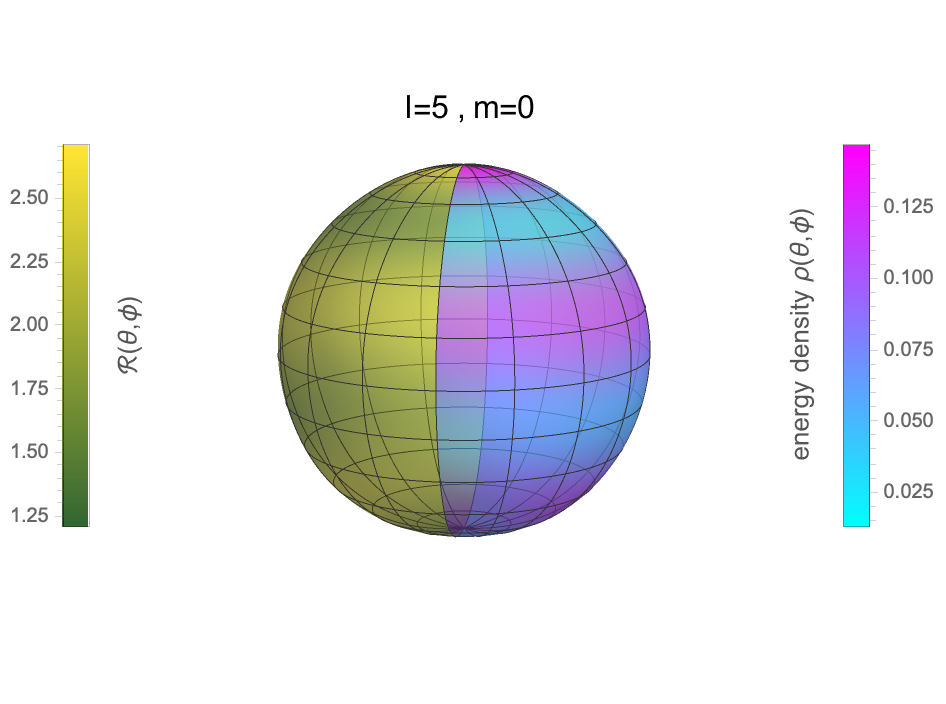}
\vspace{-1.0cm}
\caption{Geometric polarization of a conformal plasma living on a topological 2-sphere as determined by polar TLNs with $l=2,3,4,5$. All cases shown are axisymmetric, $m=0$, and adopted the value $M/L=1$ for the mass of the AdS BH dual to the plasma. The color coding on the left hemisphere represents the amount of Ricci curvature, whereas the color coding on the right hemisphere indicates the plasma energy density. The shape of the 2-spheres is fixed by isometric embeddings of the spatial part of the boundary metric in three-dimensional Euclidean space.} \label{fig:axisymmetricSpheres}
\end{figure*}
%%%%%%%%%%%%%%%%%%%%%%%%%%%

It is possible to provide a visual description of this geometric polarization effect for the case of polar sector perturbations. We do this as follows.
First, we take a constant time slice of the boundary metric, \emph{including the tidal perturbations}. This defines a line element on a 2-sphere. Generically, it will not be round. However, the axial perturbations are such that this procedure would yield a round $S^2$ and this is why we chose not to tackle the axial sector with this (less-appealing, in this case) visualization technique. Then we represent this topological $S^2$ by an isometric embedding in $\mathbb{R}^3$. I.e., we find a two-dimensional surface (homeomorphic to a 2-sphere) whose metric induced by the Euclidean metric in $\mathbb{R}^3$ precisely matches the line element we computed previously for the boundary CFT space.\footnote{See~\cite{Bondarescu:2001jf} for a very similar study, although in that case the goal was to represent the event/apparent horizon --- obtained from numerical simulations in asymptotically flat spacetimes --- through an isometric embedding in $\mathbb{R}^3$.} This can be done for axisymmetric modes ($m=0$) but non-axisymmetric modes ($m\neq0$) are more challenging. 
Finally, we color the surface of this deformed sphere according to the intensity of some component of the boundary stress-energy tensor.

Such a visualization is presented in Fig.~\ref{fig:axisymmetricSpheres} for the energy density, $T_{00}$, but one could equally do it for other components of the stress-energy tensor. According to Eqs.~(\ref{eq:stress_tot}-\ref{eq:stress_polar}) and~\eqref{eq:xi_1}, the total energy density depends on the tidal field, which is controlled by $C^{\rm p}_+$. Therefore, we must pick a value of $C^{\rm p}_+$ in order to produce these plots. In principle, $C^{\rm p}_+$ must be small in order for our linear analysis to be valid. However, to generate visible deformations of the sphere we would like $C^{\rm p}_+\lsim 1$. We have chosen $C^{\rm p}_+\in [0.2,0.5]$. 
%Larger $C^{\rm p}_+$ can produce problems. In some parts of the deformed sphere the energy density becomes \emph{negative}, which is just an artifact of our choice of tidal fields so strong that our linear perturbative analysis is invalid. Moreover, when $C^{\rm p}_+$ is sufficiently large it may happen that the deformed sphere can no longer be embedded in $\mathbb{R}^3$.

Inspection of Fig.~\ref{fig:axisymmetricSpheres} reveals that the energy density and the local curvature of the deformed sphere are correlated or anti-correlated depending on the multipole number $l$. This effect, which can be traced back to expression~\eqref{eq:xi_1}, is independent of the choice of the strength of the tidal field, $C^{\rm p}_+$, but it does depend on the ratio $M/L$ considered. For the case $M/L=1$ displayed in Fig.~\ref{fig:axisymmetricSpheres} the phenomenon is such that the energy density is largest where the sphere is less curved when $l=2,3$, whereas for all higher multipoles the energy density is largest where the sphere is more curved. We remark that the increase of $M/L$ makes this anti-correlation effect extend to higher values of $l$. For instance, for $M/L=10$ the energy density and the local curvature are anti-correlated for $l=2,\dots,6$ and correlated for all other multipoles. Conversely, for $M/L=0.1$ we only find anti-correlation when $l=2$, and for $M/L=0.01$ the anti-correlation effect is entirely absent.

%%%%%%%%%%%%%%%%%%%%%%%%%%%
%%%%%%%  SECTION 6  %%%%%%%
%%%%%%%%%%%%%%%%%%%%%%%%%%%
\section{Conclusions and discussion \label{s:discussion}}
%%%%%%%%%%%%%%%%%%%%%%%%%%%

In conclusion, this study has extended our understanding of TLNs for AdS BHs and their holographic dual plasma systems. Besides the explicit calculation of the TLNs in global AdS, in Secs.~\ref{s:LoveSAdS4} and~\ref{s:LoveKI}, we have also given a precise map between the RW and the KI formalism. 
The relative difference between our RW numerical results and those using the KI approach, converted using Eq.~\eqref{polarRWtoKI} and Eq.~\eqref{axialRWtoKI} is sub-percent for all $2\leq l\leq 50$. This can be verified explicitly with the values given in Tables~\ref{tab} and~\ref{tab2}.

To obtain our numerical results for the TLNs we employed two distinct routines:\ a direct integration and a shooting method. The direct integration routine is typically faster, but for large mass compared to the AdS length it requires a careful convergence analysis to produce reliable results for the smallest multipoles ($l\lesssim 5$). On the other hand, the shooting routine is often slower but produces more accurate results, especially for small $l$. However, large $l$ combined with small BH masses are challenging to deal with using this routine.
The two routines therefore complement each other and have very broad regimes of overlapping applicability, in which case the results obtained are in excellent agreement.

On the holographic side, while previous research~\cite{Emparan:2017qxd} has exclusively focused on black branes with plane wave perturbations using the KS formalism, here we have addressed the gap in the literature by calculating TLNs for the four-dimensional Schwarzschild solution in global AdS, corresponding to a conformal plasma on $S^2$. While performing the calculations we also clarified the connection between the KS and the KI formalisms in order to compare the two results. Moreover, we obtained an explicit visualization of the geometric polarization effect, illustrating how the dual plasma energy density is (anti-)correlated with the curvature of the deformed $S^2$ (its ambient space) depending on the mass $M$ of the bulk BH and the multipole $l$ of the perturbation considered.

A natural extension of the calculations presented here would be to consider the Kerr-AdS case, as well as exploring higher-dimensional settings, which may also be of interest. In that respect, it is worth noting that our results conform to the trace anomaly:\ in the 4D bulk case we exclusively consider, ${T_\mu}^\mu=0$, as would be true for any even $D>2$, but not for odd $D$. For instance, in $D=5$ there is a Casimir energy --- pure AdS$_5$ has a non-trivial mass. It could be interesting to investigate if the presence of a Casimir energy has any effect on the system linear response.
Additionally, a potential extension to AdS of the recently discussed dynamical TLNs~\cite{Saketh:2023bul, Perry:2023wmm, Chakraborty:2023zed} could be worth investigating.

In the case of a null cosmological constant the vanishing of BH TLNs has been understood as a consequence of recently uncovered hidden symmetries~\cite{Hui:2021vcv}. The presence of such a symmetry in a Schwarzschild background allows one to analytically obtain the whole tower of (vanishing) TLNs from just the $l=2$ mode. Since we have seen that in the Schwarzschild-AdS$_4$ case the TLNs are not zero, this means that a similar ladder structure is destroyed or, if at all present, at least modified. The existence of these symmetries seems to be inherently connected to the special properties of hypergeometric functions, which are the general form of the solutions to the perturbation equations for a Schwarzschild background. Still, it would be interesting to investigate if there are some residual symmetries that can be used to relate TLNs of Schwarzschild-AdS$_4$ at different levels (for example, with different $l$).
The so-called ``Love symmetry'', worked out in Ref.~\cite{Charalambous:2022rre},  might have a better chance of working in our context than the previously-mentioned ladder structure.
Yet another possible direction might be to exploit the recursive structure revealed in~\cite{Aminov:2023jve} to address linear perturbations of the four-dimensional Schwarzschild-AdS BH.

At any rate, TLNs have prime importance for both gravitational wave physics and for holography, explaining why they have been under intense scrutiny in recent years and will remain in the spotlight.

%%%%%%%%%%%%%%%%%%%%%%%%%%
\acknowledgments
%%%%%%%%%%%%%%%%%%%%%%%%%%
We thank Vitor Cardoso and Roberto Emparan for interesting and useful discussions.
EF acknowledges funding support from the MUR PRIN (Grant 2020-KR4KN2 ``String Theory as a bridge between Gauge Theories and Quantum Gravity''), FARE (GW-NEXT, CUP:\ B84I20000100001) programmes, and from EU Horizon 2020 Research and Innovation Programme under the Marie Sk\l{}odowska-Curie Grant Agreement no.\ 101007855.
AMF acknowledges the support of the MICINN grants PID2019-105614GB-C22, AGAUR grant 2017-SGR 754, PID2022-136224NB-C22 funded by MCIN/AEI/ 10.13039/501100011033/FEDER, UE, and State Research Agency of MICINN through the ``Unit of Excellence Maria de Maeztu 2020--2023'' award to the Institute of Cosmos Sciences (CEX2019-000918-M) and from EU Horizon 2020 Research and Innovation Programme under the Marie Sk\l{}odowska-Curie Grant Agreement no.\ 101149470.
JVR acknowledge support from FCT under Project no.\ 2022.08368.PTDC and Project no.\ CERN/FIS-PAR/0023/2019.

\pagebreak

\appendix

%%%%%%%%%%%%%%%%%%%%%%%%%%%
%%%%%%  Appendix A  %%%%%%%
%%%%%%%%%%%%%%%%%%%%%%%%%%%
\section{Asymptotic expansions in the Regge-Wheeler formalism\label{a:asymptotics}}
%%%%%%%%%%%%%%%%%%%%%%%%%%%

In the AdS case, i.e., when the cosmological constant is strictly negative, the asymptotic expansions of the polar and axial perturbations in the RW gauge can be written as
\be
H = \, \frac{C^\text{p}_-}{r} \sum_{i=0}^\infty \frac{p_i^-}{r^i} \; + \;  \frac{C^\text{p}_+}{r^2} \sum_{i=0}^\infty \frac{p_i^+}{r^i}\,,
\qquad\qquad
h_0 = C^\text{a}_+ \, r^2 \sum_{i=0}^\infty \frac{a_i^+}{r^i} \; + \; \frac{C^\text{a}_-}{r} \sum_{i=0}^\infty \frac{a_i^-}{r^i}\,.
\ee
The coefficients $p_i^+$, $p_i^-$, $a_i^+$ and $a_i^-$ can be computed by inserting the above expressions into the corresponding differential equations and solving them order by order in powers of $r^{-1}$. Obviously, the zeroth order coefficients can be normalized to unity. For the remaining terms, we find for the polar sector
\begin{subequations}\be
p_1^- &= 0\,,
& p_1^+ &= 0\,,\\
p_2^- &= - \frac{3(l^2+l-4)}{2\Lambda}\,,
& p_2^+ &= - \frac{(l-2)(l+3)}{2\Lambda}\,,\\
p_3^- &= -\frac{9M}{\Lambda}\,,
& p_3^+ &= -\frac{6M}{\Lambda}\,,\\
p_4^- &= \frac{3(l-2)(l+3)\left( l^2+l-8 \right)}{8\Lambda^2}\,,
& p_4^+ &= \frac{3(l-3)(l+4)\left( l^2+l-10 \right)}{40\Lambda^2}\,,\\
p_5^- &= \frac{9M(l^2+l-7)}{\Lambda^2}\,,
& p_5^+ &= \frac{3M(l-3)(l+4)}{\Lambda^2}\,,
\ee\end{subequations}
and for the axial sector
\begin{subequations}\be
a_1^+ &= 0\,,
& a_1^- &= 0\,,\\
a_2^+ &= \frac{3(l-1)(l+2)}{2\Lambda}\,,
& a_2^- &= - \frac{3(l-1)(l+2)}{10\Lambda}\,,\\
a_3^+ &= 0\,,
& a_3^- &= 0\,,\\
a_4^+ &= - \frac{9(l-1)l(l+1)(l+2)}{8\Lambda^2}\,,
& a_4^- &= \frac{9(l-3)(l-1)(l+2)(l+4)}{280\Lambda^2}\,,\\
a_5^+ &= \frac{9M (l-1)(l+2)}{5 \Lambda^2}\,,
& a_5^- &= \frac{9M(l-1)(l+2)}{20\Lambda^2}\,.
\ee\end{subequations}

%%%%%%%%%%%%%%%%%%%%%%%%%%%
%%%%%%  Appendix B  %%%%%%%
%%%%%%%%%%%%%%%%%%%%%%%%%%%
\section{Series expansions around the horizon in the Regge-Wheeler formalism\label{a:horizon}}
%%%%%%%%%%%%%%%%%%%%%%%%%%%

Imposing regularity of the perturbations at the horizon implies that both the polar and axial master variables can be expanded in Taylor series around the horizon,
\be
H = \, \sum_{i=0}^\infty \alpha_i\left(\frac{r-r_h}{r_h}\right)^i \,,
\qquad\qquad
h_0 = \, \sum_{i=0}^\infty \beta_i\left(\frac{r-r_h}{r_h}\right)^i\,.
\ee
The coefficients $\alpha_i$ and $\beta_i$ can be computed by inserting the above expressions into the corresponding differential equations for $H$ and $h_0$ and solving them order by order in powers of $(r-r_h)$. Note that one must express either $\Lambda$ or $M$ as a function of $r_h$ in order to perform these expansions, since the three quantities obey the constraint $f(r_h)=0$; here we opted by eliminating $\Lambda$. It turns out that the zeroth order coefficient identically vanishes, while all other coefficients are proportional to the first order ones. (We may, without loss of generality normalize the first ones to unity, keeping in mind that the differential equations satisfied by $H$ and $h_0$ are homogeneous, but here we keep that parameter free.) In the polar sector we find
\begin{subequations}\be
\alpha_0 &= 0\,,\\
\alpha_2 &= \frac{\left(l^2+l+9\right)r_h - 24M}{6(3M-r_h)} \alpha_1\,,\\
\alpha_3 &= \frac{\left(l^4 + 2l^3 + 39l^2 + 38l + 168\right) r_h^2 - 96\left(l^2+l+9\right) M r_h + 1152M^2}{96(3M - r_h)^2} \alpha_1\,,\\
\alpha_4 &= \frac{\alpha_1}{2880\, (3M-r_h)^3} \Big[ \left(l^6 + 3l^5 + 100l^4 + 195l^3 + 1939l^2 + 1842l + 5400\right) r_h^3 \nn\\
&\phantom{=} - 240 \left(l^4 + 2l^3 + 39l^2 + 38l + 168\right) M r_h^2 + 1440 \left(8l^2 + 8l + 73\right) M^2 r_h - 95040 M^3\Big] \,,\\
\alpha_5 &= \frac{\alpha_1}{(r_h-3 M)^4} \Big[ 90 M^4 -\frac{3}{2} \left(9 l^2+9 l+86\right) M^3 r_h +\frac{5 l^4+10 l^3+196 l^2+191 l+864}{12} M^2 r_h^2\nn\\
&\phantom{=} -\frac{l^6+3 l^5+100 l^4+195 l^3+1939 l^2+1842 l+5400}{288}  M r_h^3\nn\\
&\phantom{=} +\frac{l^8+4 l^7+202 l^6+592 l^5+10129 l^4+19276 l^3+130788 l^2+121248 l+267840}{138240} r_h^4\Big] \,.
\ee\end{subequations}

In the axial sector we obtain
\begin{subequations}\be
\beta_0 &= 0\,,\\
\beta_2 &= \frac{(l+2)(l-1)r_h}{4(3M - r_h)} \beta_1\,,\\
\beta_3 &= \frac{ l \left(l^3+2 l^2+7 l+6\right) r_h^2 -24 \left(l^2+l+2\right) M r_h + 144 M^2}{48 (3M-r_h)^2} \beta_1\,,\\
\beta_4 &= \frac{\beta_1}{1152 (3M-r_h)^3}  \Big[ l \left(l^5 + 3l^4 + 37l^3 + 69l^2 + 154l + 120\right) r_h^3\nn\\
&\phantom{=} - 96 \left(l^4 + 2l^3 + 11l^2 + 10l + 12\right) M r_h^2 + 1728 \left(l^2+l+4\right) M^2 r_h - 10368M^3 \Big]\,,\\
\beta_5 &= \frac{\beta_1}{46080\, (3M-r_h)^4} \Big[ 
l \left(l^7+4l^6+98l^5+280l^4+1889l^3+3316 l^2+5692l+4080\right) r_h^4 -\nn\\
&\phantom{=} - 240 \left(l^6+3 l^5+41 l^4+77 l^3+246 l^2+208 l+192\right) M r_h^3 + 1244160 M^4\nn\\
&\phantom{=}+ 12960 \left(l^4+2 l^3+15 l^2+14 l+32\right) M^2 r_h^2 - 207360 \left(l^2+l+6\right) M^3 r_h  \Big] \,.
\ee\end{subequations}

%%%%%%%%%%%%%%%%%%%%%%%%%%%
%%%%%%  Appendix C  %%%%%%%
%%%%%%%%%%%%%%%%%%%%%%%%%%%
\section{Relation between Kodama-Ishibashi and Regge-Wheeler master equations\label{a:master_relations}}
%%%%%%%%%%%%%%%%%%%%%%%%%%%

Equation~\eqref{eq:MappingScalar}, relating the KI master variable $\Phi_S$ to the polar sector variable $H$ in the RW formalism, schematically maps the left hand side of~\eqref{eq:MasterScalar} to
\be
&\frac{2 \; \text{LHS}(\ref{eq:Polar-AdS})}{r^3 \left[9\mathfrak{m} (\mathfrak{m}+2) (\mathfrak{m}+3 x)^2 (x+y-1)\right]} 
\Big\{ \mathfrak{m}^2 \Big[(\mathfrak{m}+6)-(\mathfrak{m}+10)y\Big]\nn\\
&\qquad+\mathfrak{m} \Big[12(\mathfrak{m}-1)y -(\mathfrak{m}+2) (\mathfrak{m}+8) -18y^2 +22\Big]x
+3\Big[(\mathfrak{m}-7)\mathfrak{m}+3(5\mathfrak{m}-2)y-6y^2\Big]x^2\nn\\
&\qquad+3\Big[3(\mathfrak{m}-2)+24y\Big]x^3 + 9x^4 \Big\}
+ \frac{2 \left[x (6 y-3 \mathfrak{m})+2 \mathfrak{m}-3 x^2\right]}{r^2 [9\mathfrak{m} (\mathfrak{m}+2) (\mathfrak{m}+3 x)]}\,\frac{d}{dr}\text{LHS}(\ref{eq:Polar-AdS})\,. 
\ee

A corresponding relation also exists for the axial sector, for which a more compact form can be found. Using Eq.~\eqref{eq:MasterVector}, which relates the RW variable $h_0$ to the vector master variable $\Phi_V$ in the KI formalism, the left hand side of~\eqref{eq:Axial-AdS} gets mapped to
\be
\frac{6r^2 f(r)-2r^3f'(r)}{f(r)^2}\,\text{LHS}(\text{\ref{eq:MasterVector}}) + \frac{2r^3}{f(r)} \,\frac{d}{dr}\text{LHS}(\text{\ref{eq:MasterVector}})\,.
\ee

%%%%%%%%%%%%%%%%%%%%%%%%%%%
%%%%%%  Appendix D  %%%%%%%
%%%%%%%%%%%%%%%%%%%%%%%%%%%
\section{Asymptotic expansions in the Kodama-Ishibashi formalism \label{a:asymptotics_KI}}
%%%%%%%%%%%%%%%%%%%%%%%%%%%

In the AdS case, i.e., when the cosmological constant is strictly negative, the asymptotic expansions of the polar and axial perturbations in the KI formalism can be written as
\be
\Phi_S = \, D^\text{s}_+ \sum_{i=0}^\infty \frac{s_i^+}{r^i} \; + \;  \frac{D^\text{s}_-}{r} \sum_{i=0}^\infty \frac{s_i^-}{r^i}\,,
\qquad\qquad
\Phi_V = D^\text{v}_+ \sum_{i=0}^\infty \frac{v_i^+}{r^i} \; + \; \frac{D^\text{v}_-}{r} \sum_{i=0}^\infty \frac{v_i^-}{r^i}\,.
\ee
The coefficients $s_i^+$, $s_i^-$, $v_i^+$ and $v_i^-$ can be computed by inserting the above expressions into the corresponding differential equations and solving them order by order in powers of $r^{-1}$. Obviously, the zeroth order coefficients can be normalized to unity. For the remaining terms, we find, for the scalar sector [recall that $\mathfrak{m}=(l-1)(l+2)$]
\begin{comment}
\begin{subequations}\be
s_1^+ &= 0\,,
& s_1^- &= 0\,,\\
%
s_2^+ &=  \frac{36M^2}{\mathfrak{m}^2}-\frac{3(\mathfrak{m}+2)}{2\Lambda}\,,
& s_2^- &=  \frac{12 M^2}{\mathfrak{m}^2} - \frac{\mathfrak{m}}{2 \Lambda}\,,\\
%
s_3^+ &=  \frac{3 M}{\Lambda} \left( -1 + \frac{2(\mathfrak{m}+2)}{\mathfrak{m}} - \frac{48 M^2 \Lambda}{\mathfrak{m}^3} \right),
& s_3^- &= \frac{6 M}{\mathfrak{m}\Lambda} \left(1 - \frac{12 M^2 \Lambda}{\mathfrak{m}^2} \right),\\
%
s_4^+ &= 3 M^4\left( \frac{288}{\mathfrak{m}^4}-\frac{12 \left(\mathfrak{m}+1\right)}{ M^2 \Lambda\, \mathfrak{m}^2}+\frac{(\mathfrak{m}+2)(\mathfrak{m}-4)}{8 M^2 \Lambda ^2}\right),
& s_4^- &= 3 M^4 \left(\frac{144}{\mathfrak{m}^4}  - \frac{24 (\mathfrak{m}+1) }{5 M^2 \Lambda\, \mathfrak{m}^2} +\frac{(\mathfrak{m}-10)\mathfrak{m}}{40 M^4 \Lambda^2} \right),\\
%s_5^+ &=& \frac{9M \left(24 \Lambda (3m+8) m^2 M^2+\left(m^2+m+4\right) m^4-2880 \Lambda^2 M^4\right)}{5\Lambda^2 m^5} \,,
%\qquad\qquad\qquad\qquad\\
s_5^+ &= \frac{9 M^5}{5} \left( \frac{24 (3\mathfrak{m}+8)}{ M^2 \Lambda \mathfrak{m}^3}+\frac{\mathfrak{m}^2+\mathfrak{m}+4}{M^4 \Lambda^2 \mathfrak{m}}- \frac{2880}{ \mathfrak{m}^5}\right),
& s_5^- &= 3M^5 \left(\frac{12(7\mathfrak{m}+12)}{5M^2\Lambda\, \mathfrak{m}^3} + \frac{2\mathfrak{m}^2-3\mathfrak{m}+18}{5M^4 \Lambda ^2 \mathfrak{m}}-\frac{864}{\mathfrak{m}^5}\right),
\ee\end{subequations}
\end{comment}
\begin{subequations}\be
s_1^+ &= 0\,,\quad 
s_2^+ =  \frac{36M^2}{\mathfrak{m}^2}-\frac{3(\mathfrak{m}+2)}{2\Lambda}\,,\quad
s_3^+ =  \frac{3 M}{\Lambda} \left( -1 + \frac{2(\mathfrak{m}+2)}{\mathfrak{m}} - \frac{48 M^2 \Lambda}{\mathfrak{m}^3} \right),\\
s_4^+ &= 3 M^4\left( \frac{288}{\mathfrak{m}^4}-\frac{12 \left(\mathfrak{m}+1\right)}{ M^2 \Lambda\, \mathfrak{m}^2}+\frac{(\mathfrak{m}+2)(\mathfrak{m}-4)}{8 M^2 \Lambda ^2}\right),\\
s_5^+ &= \frac{9 M^5}{5} \left( \frac{24 (3\mathfrak{m}+8)}{ M^2 \Lambda \mathfrak{m}^3}+\frac{\mathfrak{m}^2+\mathfrak{m}+4}{M^4 \Lambda^2 \mathfrak{m}}- \frac{2880}{ \mathfrak{m}^5}\right),\\
s_1^- &= 0\,,\quad
s_2^- =  \frac{12 M^2}{\mathfrak{m}^2} - \frac{\mathfrak{m}}{2 \Lambda}\,,\quad
s_3^- = \frac{6 M}{\mathfrak{m}\Lambda} \left(1 - \frac{12 M^2 \Lambda}{\mathfrak{m}^2} \right),\\
s_4^- &= 3 M^4 \left(\frac{144}{\mathfrak{m}^4}  - \frac{24 (\mathfrak{m}+1) }{5 M^2 \Lambda\, \mathfrak{m}^2} +\frac{(\mathfrak{m}-10)\mathfrak{m}}{40 M^4 \Lambda^2} \right),\\
s_5^- &= 3M^5 \left(\frac{12(7\mathfrak{m}+12)}{5M^2\Lambda\, \mathfrak{m}^3} + \frac{2\mathfrak{m}^2-3\mathfrak{m}+18}{5M^4 \Lambda ^2 \mathfrak{m}}-\frac{864}{\mathfrak{m}^5}\right),
\ee\end{subequations}
and for the vector sector
\begin{subequations}\be
v_1^+ &= 0\,, 
& v_1^- &= 0\,,\\
v_2^+ &= -\frac{3l(l+1)}{2\Lambda}\,,
& v_2^- &= -\frac{(l-1)(l+2)}{2\Lambda}\,,\\
v_3^+ &= \frac{3M}{\Lambda}\,,
& v_3^- &= 0\,,\\
v_4^+ &= \frac{3(l-2)l(l+1)(l+3)}{8\Lambda^2}\,, 
& v_4^- &= \frac{3(l-3)(l-1)(l+2)(l+4)}{40\Lambda^2}\,,\\
v_5^+ &= \frac{9M \left(l^2+l+3\right)}{5\Lambda^2}\,,
& v_5^- &= \frac{6M (l-1)(l+2)}{5\Lambda^2}\,.
\ee\end{subequations}

%%%%%%%%%%%%%%%%%%%%%%%%%%%
%%%%%%  Appendix E  %%%%%%%
%%%%%%%%%%%%%%%%%%%%%%%%%%%
\section{Series expansions around the horizon in the Kodama-Ishibashi formalism\label{a:horizonKI}}
%%%%%%%%%%%%%%%%%%%%%%%%%%%

Imposing regularity of the perturbations at the horizon implies that both the scalar and vector master variables can be expanded in Taylor series around the horizon,
\be
\Phi_S = \, \sum_{i=0}^\infty \sigma_i\left(\frac{r-r_h}{r_h}\right)^i \,,
\qquad\qquad
\Phi_V = \, \sum_{i=0}^\infty \nu_i\left(\frac{r-r_h}{r_h}\right)^i\,.
\ee
Similar to what was done in Appendix~\ref{a:horizon}, the coefficients $\sigma_i$ and $\nu_i$ can be computed by inserting the above expressions into the corresponding differential equations for $\Phi_S$ and $\Phi_V$ and solving them order by order in powers of $(r-r_h)$. Once again, we express $\Lambda$ as a function of $r_h$ and $M$ in order to perform these expansions. In contrast to what happened in Appendix~\ref{a:horizon},  now the zeroth order coefficient does not vanish, all other coefficients being proportional to the them. (We may, without loss of generality normalize the zeroth order coefficients to unity, keeping in mind that the differential equations satisfied by $\Phi_S$ and $\Phi_V$ are homogeneous, but here we keep that parameter free.) In the scalar sector we find:
\begin{subequations}\be
\sigma_1 &= \frac{\sigma_0}{6M+\mathfrak{m}\, r_h} \left[ 
6M + \frac{\mathfrak{m}(\mathfrak{m}+2) r_h^2}{2(3M-r_h)}\right]\,,\\
\sigma_2 &= \frac{\sigma_0 \mathfrak{m}\, r_h }{(6M+\mathfrak{m}\, r_h)^2} \left[-6M - \frac{\mathfrak{m}(\mathfrak{m}+2) r_h^2}{4 (3M-r_h)} + \frac{\mathfrak{m}(\mathfrak{m}+2)(\mathfrak{m}+4) r_h^3+36 (\mathfrak{m}+2) M^2 r_h}{16 (3 M-r_h)^2}\right]\,,\\
\sigma_3 &= \frac{\sigma_0\mathfrak{m}(\mathfrak{m}+2) r_h^2}{6 (6M+\mathfrak{m} r_h)^3}\left[\frac{(29 \mathfrak{m}-14)M}{\mathfrak{m}+2} -\frac{(13 \mathfrak{m}+11) r_h}{3} +\frac{\left(\mathfrak{m}^2-14 \mathfrak{m}-8\right) r_h^2}{2 (3 M-r_h)}\right.\nn\\
&\phantom{=} \left.-\frac{(5 \mathfrak{m}-2) (\mathfrak{m}+2)^2 r_h^3}{12 (3 M-r_h)^2} +\frac{(\mathfrak{m}+6) (\mathfrak{m}+2)^3 r_h^4}{48 (3 M-r_h)^3}\right].
\ee\end{subequations}
In the vector sector we obtain
\begin{subequations}\be
\nu_1 &= -\nu_0 \left[1-\frac{\mathfrak{m} r_h}{2 (3 M-r_h)}\right],\\
\nu_2 &= \nu_0 \left[1 - \frac{\mathfrak{m} r_h}{2 (3 M-r_h)} + \frac{\mathfrak{m} (\mathfrak{m}+2) r_h^2}{16 (3 M-r_h)^2}\right],\\
\nu_3 &= -\nu_0 \left[1 - \frac{\mathfrak{m} r_h}{2 (3 M-r_h)} + \frac{13 \mathfrak{m} (\mathfrak{m}+2) r_h^2}{144 (3 M-r_h)^2} -\frac{\mathfrak{m} (\mathfrak{m}+2) (\mathfrak{m}+6) r_h^3}{288 (3 M-r_h)^3}\right].
\ee\end{subequations}

Formulas for the higher order $\sigma_i$ and $\nu_i$ can be straightforwardly obtained, but we refrain from displaying them because they are lengthy and uninformative.

%%%%%%%%%%%%%%%%%%%%%%%%%%%
%%%%%%  Appendix F  %%%%%%%
%%%%%%%%%%%%%%%%%%%%%%%%%%%
\section{Solutions of the static Kodama-Ishibashi master equations on the Schwarzschild background\label{a:LoveKISchwarz}}
%%%%%%%%%%%%%%%%%%%%%%%%%%%

In this appendix we provide analytical solutions to Eqs.~\eqref{eq:MasterScalar} and \eqref{eq:MasterVector} when the background is a Schwarzschild BH, i.e., with $f(r)=1-2M/r$.

In both sectors, there are two independent solutions that behave asymptotically as $r^{l+1}$ and $1/r^l$, i.e.
\be
\Phi_S = s_1\,\Phi_S^{(1)} + s_2\,\Phi_S^{(2)}\,,
\quad\text{and}\quad
\Phi_V = v_1\,\Phi_V^{(1)} + v_2\,\Phi_V^{(2)}\,,
\ee
where $s_{1,2}$ and $v_{1,2}$ are integration constants.

In the scalar sector we look for a growing solution in the form
\be\label{PhiSsol1}
\Phi_S^{(1)} = \frac{M \sum_{i=0}^{l+2} \alpha_i \left(\frac{r}{M}\right)^i}{6M + (l-1)(l+2)r}\,.
\ee
By substituting in Eq.~\eqref{eq:MasterScalar} we find a three-term recurrence relation for the coefficients
\be
12 (i-1)^2\,\alpha_i
+ 2(i-2)\left[(i-4)l(l+1)-5i+11\right]\alpha_{i-1}&\nn\\
+ (l-1)(l+2)(l+4-i)(l-3+i)\,\alpha_{i-2} &= 0\,,
\quad\text{for $i\geq2$}\,,
\ee
with $\alpha_0 = 0\,,\alpha_1 = 1$.
We notice that this solution is regular at the horizon, but it is not a polynomial and for large $r$ it also contains $1/r^l$ terms.

The other decaying solution can be found by considering
\be\label{PhiSsol2}
\Phi_S^{(2)} = \frac{M \sum_{i=0}^{l+2} \beta_i \left(\frac{r}{M}\right)^i}{6M + (l-1)(l+2)r}
+ \Sigma\,\Phi_S^{(1)}\log\left(1-\frac{2M}{r}\right),
\ee
with the constant $\Sigma$ to be determined by requiring that $\Phi_S^{(2)}\sim 1/r^l$ as $r\to\infty$.
Substituting in Eq.~\eqref{eq:MasterScalar} we find a recurrence relation that also depends on $\alpha_i$,
\be
12(i-1)^2\,\beta_i
+ 2(i-2)\left[(i-4)l(l+1)-5i+11\right]\beta_{i-1}&\nn\\
+ (l-1)(l+2)(l+4-i)(l-3+i)\,\beta_{i-2}&\nn\\
- 4\Sigma\left[6(i-1)\,\alpha_i + (i-3)(l-1)(l+2)\,\alpha_{i-1}\right]
&= 0\,,
\quad\text{for $i\geq2$}\,,
\ee
with $\beta_0 = 0\,,\beta_1 = 1$.
By construction, $\Phi_S^{(2)}$ contains logarithmic terms and it is not regular at the horizon $r=2M$.

Similarly, in the vector sector, we look for a growing solution as
\be
\Phi_V^{(1)} = \left(\frac{r}{2M}\right)^{l+1}\sum_{i=0}^\infty \alpha_i \left(\frac{2M}{r}\right)^i\,.
\ee
In this case, the coefficients satisfy a simple recurrence relation that can be solved, and by resumming the infinite series we get a general expression in terms of the hypergeometric function ${}_2F_1(a,b;c;x)$
\be
\Phi_V^{(1)}
=\left(\frac{r}{2 M}\right)^3 \, {}_2F_1\left(2-l,l+3;5;\frac{r}{2M}\right).
\ee
Notice that for any given value of $l$ this hypergeometric contains a finite number of terms and reduces to a polynomial of degree $l-2$, while the whole solution is a polynomial of order $l+1$.

For the other independent solution, we start with
\be\label{PhiVsol1}
\Phi_V^{(2)} = \left(\frac{2M}{r}\right)^l\sum_{i=0}^\infty \beta_i \left(\frac{2M}{r}\right)^i\,,
\ee
and even in this case the recurrence equation can be solved and the series resummed, yielding
\be\label{PhiVsol2}
\Phi_V^{(2)} = \left(\frac{2M}{r}\right)^l \, {}_2F_1\left(l-1,l+3;2 l+2;\frac{2M}{r}\right).
\ee
This solution is a rational function whose numerator contains terms proportional to $\log\left(1-2M/r\right)$ and hence not regular at the horizon.

We remind that in the RW gauge the perturbation equations~\eqref{eq:Hpert} and \eqref{eq:h0pert} on a Schwarzschild background have general solutions
\be
H(r) &= a_1\,P_l^2\left(\frac{r}{M}-1\right) + a_2\,Q_l^2\left(\frac{r}{M}-1\right),\label{solH-RW}\\
h_0(r) &= b_1 \left(\frac{r}{2M}\right)^2 {}_2F_1\left(1-l,l+2;4;\frac{r}{2 M}\right) + b_2 \left(\frac{2 M}{r}\right)^l \, _2F_1\left(l-1,l+2;2 l+2;\frac{2 M}{r}\right),\label{solh0-RW}
\ee
where $P_l^2$ and $Q_l^2$ are the associated Legendre functions of first and second kind, of degree $l$ and order $2$.
$a_1$, $a_2$, $b_1$ and $b_2$ are arbitrary integration constants. The solutions associated to $a_2$ and $b_2$ contain terms proportional to $\log(1-2M/r)$, and the requirement that the solutions \eqref{solH-RW} and \eqref{solh0-RW} are everywhere regular (including on the event horizon) implies $a_2=b_2=0$. This, in turn, means that the TLNs for a Schwarzschild BH are identically zero.

Now, with the relation \eqref{eq:MappingScalar} one finds that the solution proportional to $a_1$, respectively $a_2$, in Eq.~\eqref{solH-RW} gets mapped (modulo constant factors) in Eq.~\eqref{PhiSsol1}, respectively in Eq.~\eqref{PhiSsol2}.
Likewise, with the relation \eqref{eq:MappingVector}, the solution proportional to $b_1$, respectively $b_2$, in Eq.~\eqref{solh0-RW} gets mapped in Eq.~\eqref{PhiVsol1}, respectively in Eq.~\eqref{PhiVsol2}.
Since $a_2=b_2=0$ because of regularity, $s_2$ and $v_2$ must vanish as well, and as a consequence, TLNs for a Schwarzschild BH are zero in the KI formalism as well.

%%%%%%%%%%%%%%%%%%%%%%%%%%%
%%%%%%  Appendix G  %%%%%%%
%%%%%%%%%%%%%%%%%%%%%%%%%%%
\section{Relation with tidal Love numbers of black branes using the Kovtun-Starinets formalism\label{s:LoveKS}}
%%%%%%%%%%%%%%%%%%%%%%%%%%%

Taking the eikonal limit (i.e., $l\to\infty$), it is possible to compare our results with the findings of Ref.~\cite{Emparan:2017qxd}. This ultraviolet regime probes the metric only near the boundary of AdS and is therefore insensitive to whether the spacetime has planar or spherical symmetry (and also to the presence or not of a BH in the bulk). 
To explicitly make the connection, we have to relate the multipole number $l$, appropriate for the spherical harmonic decomposition of perturbations of BHs in global AdS, with the wavenumber $k$ of plane waves used in the mode decomposition of perturbations of black branes in AdS.

Recall that the Laplacian operator on maximally symmetric non-compact manifolds has a continuous non-negative spectrum. In particular, when applied to plane waves of the form considered above, the eigenvalue is $-k^2$ as considered for black branes in AdS.  On the other hand, for BHs in global AdS, the internal manifold ---in the language of~\cite{Ishibashi:2011ws}--- is a compact 2-sphere, and the spectrum is discrete with the usual dimensionless spherical harmonic eigenvalues $-l(l+1)$ with $l \in \mathbb{N}$.
In the eikonal limit $l \rightarrow \infty$, the eigenvalues match if we replace $l^2 \rightarrow \hat{k}^2 \equiv (Lk)^2$.

Using the results for AdS black branes in Ref.~\cite{Morgan:2009pn, Mamani:2018qzl} with $d=4$, the general solution to the KI scalar master equation for black branes behaves asymptotically (i.e., for small holographic radial coordinate $u\equiv r_h/r$) as
\be \label{eq:black_brane_KI}
\Phi _S(u) &= D_+^\text{s} \left[1 + \left(\frac{3 \hat{k}^2}{2} + \frac{\left(-\Lambda r_h^2\right)^3}{\hat{k}^4}\right)\frac{u^2}{-\Lambda  r_h^2} + \dots \right]\nn\\
&\phantom{=} + D_-^\text{s} \frac{u}{r_h} \left[1+ \frac{1}{3}\left(\frac{3 \hat{k}^2}{2} + \frac{\left(-\Lambda r_h^2\right)^3}{\hat{k}^4}\right) \frac{u^2}{-\Lambda  r_h^2} + \dots \right]\,.
\ee

In the KS formalism~\cite{Kovtun:2005ev}, a different master variable $Z_S$ is adopted, and it obeys the following master equation~\cite{Miranda:2008vb},
\be
Z_S''(u) + \frac{{\cal Y}_1}{u f{\cal X}} Z_S'(u) + \frac{{\cal Y}_3 + {\cal Y}_4 \mathfrak{q}^2}{f {\cal X}} Z_S(u) = 0\,,
\label{eq:KSmasterEq}
\ee
where the parameter $\mathfrak{q} \equiv \frac{L^2}{r_h} k$ represents a normalized wavenumber and $f$ is   $f(u) = 1 - u^3$, the metric function~\eqref{eq:bkgd_metric} for the black brane written in terms of the holographic radial coordinate $u$.
The other functions in Eq.~\eqref{eq:KSmasterEq}, are defined in the following way:
\be
{\cal Y}_1 = 3 u^{3} \left(3+f\right)+8 f^2\,,\quad
{\cal Y}_3 = - f'^2\,,\quad
{\cal Y}_4 = 4- u^{3}\,,\quad
{\cal X} = - \left( f+3 \right)\,.
\ee
%$\cal{X}$ is different when compared to Eq.(3.22) in Cardoso et al. There is a typo in that paper.

The mapping between the two formalisms is provided by~\cite{Mamani:2018qzl}
\be
Z_S(u) = {\cal P}_S(u) \Phi_S'(u) + {\cal Q}_S(u) \Phi_S(u)\,,
\label{eq:mapZSPhiS}
\ee
with
\be
{\cal P}_S &= - \frac{ \mathfrak{q}^2 f^2}{3 u + \mathfrak{q}^2 }\,,\\
{\cal Q}_S &= \frac{- \mathfrak{q}^2 }{ 4  \left( \mathfrak{q}^2 + 3 u  \right)^2}  \left[  \mathfrak{q}^4 u \left( u^{3} - 4 \right) - 18 \mathfrak{q}^2 u^{2} + 3 \left( u^{6} -14 u^{3} + 4 \right) \right]\,.
\ee

Plugging the asymptotic expansion~\eqref{eq:black_brane_KI} into the mapping which relates $\Phi_S$ to the KS master variable $Z_S$ we obtain the corresponding asymptotic expansion for $Z_S(u)$,
\be
Z_S(u) = &-\left( \frac{D_-^\text{s}}{r_h} + \frac{D_+^\text{s} \left(-\Lambda r_h^2\right)}{\hat{k}^2} \right) \left[ 1 - \frac{3 \hat{k}^2 u^2}{2 \left(-\Lambda r_h^2\right)} + \dots \right] \nn\\
&+ 3 \left[ \frac{D_-^\text{s}}{2 r_h} + D_+^\text{s}\left(\frac{\hat{k}^2}{-\Lambda r_h^2}\right)^2 + \frac{D_+^\text{s} \left(-\Lambda r_h^2\right)}{2 \hat{k}^2} \right] \left(u^3 + \dots \right)\,.
\label{eq:ZSu}
\ee
In the vector sector instead, in the KS formalism, the master variable $Z_V$ obeys~\cite{Miranda:2008vb}
\be
Z_V''(u) - \frac{2}{u} Z_V'(u) - \frac{ \mathfrak{q}^2 }{f} Z_V(u) = 0\,.
\ee
and the mapping between the two vector master variables is given by~\cite{Mamani:2018qzl}
\be
Z_V(u) = u^{2} f \frac{d}{du}\left[u^{-1} \Phi_V(u)\right]\,.
\ee
The asymptotic expansion of the general solution to the KI vector master equation $\Phi_V$,
\be
\Phi_V(u) = D_+^\text{v} \left[ 1+ \frac{3 \hat{k}^2}{2} \frac{u^2}{(-\Lambda r_h^2)} + \dots \right]
+ D_-^\text{v} \frac{u}{r_h} \left[ 1+ \frac{\hat{k}^2}{2} \frac{u^2}{(-\Lambda r_h^2)} + \dots \right]\,,
\ee
is transformed into the following asymptotic expansion for $Z_V$:
\be
Z_V(u) = -D_+^\text{v} \left[ 1 - \frac{3 \hat{k}^2}{2 \left(-\Lambda r_h^2\right)} u^2 + \dots \right] + D_-^\text{v} \frac{\hat{k}^2}{\left(-\Lambda r_h^2\right) r_h} \left( u^3 + \dots \right)\,.
\label{eq:ZVu}
\ee

In Ref.~\cite{Emparan:2017qxd} the coordinate
$v = (-3 r_h/\Lambda  r_h^2) u$
was used instead of $u$, in terms of which the asymptotic expansion for $Z_S$ and $Z_V$ was written as
\be
Z_S(v) &= A^{(S)} \left[1+\dots\right] + B^{(S)} \left[ v^3+\dots \right]\,,\label{eq:ZSv}\\
Z_V(v) &= A^{(V)} \left[1+\dots\right] + B^{(V)} \left[ v^3+\dots \right]\,.\label{eq:ZVv}
\ee
By comparing expressions~\eqref{eq:ZSu} and~\eqref{eq:ZSv} one can now easily relate the coefficients $\{A^{(S)}, B^{(S)}\}$ with $\{D_+^\text{s}, D_-^\text{s}\}$,
while by comparing expressions~\eqref{eq:ZVu} and~\eqref{eq:ZVv} one relates the coefficients $\{A^{(V)}, B^{(V)}\}$ with $\{D_+^\text{v}, D_-^\text{v}\}$.

The upshot is that the scalar sector TLN $\lambda_S \equiv L^3\frac{B^{(S)}}{A^{(S)}}$, as defined in Ref.~\cite{Emparan:2017qxd} based on the KS formalism, is related with the TLN ${\cal K}^S_l = \frac{1}{L} \frac{D_-^\text{s}}{D_+^\text{s}}$, computed using the KI formalism in Sec.~\ref{s:LoveKI}, as follows
\be
\lambda_S = - \frac{2(Lk)^6 + 9 (Lk)^2 \frac{r_h^3}{L^3} {\cal K}^S_l + 27 \frac{r_h^6}{L^6}}{6 \, (Lk)^2 \,{\cal K}^S_l + 18 \frac{r_h^3}{L^3}} \;
\stackrel{k\to\infty}{\longrightarrow} \; - \frac{(Lk)^4}{3{\cal K}^S_l} \,,
\label{eq:LoveScalarRelation}
\ee
where the limit is taken for large $k$.  

The vector sector TLN defined in Ref.~\cite{Emparan:2017qxd} as $\lambda_V \equiv L^3\frac{B^{(V)}}{A^{(V)}}$, can be obtained from the KI-based TLN, ${\cal K}^V_l = \frac{1}{L} \frac{D_-^\text{v}}{D_+^\text{v}}$, according to
\be
\lambda_V = -\frac{1}{3} (Lk)^2 {\cal K}^V_l\,.
\label{eq:LoveVectorRelation}
\ee

In the eikonal limit $l\to\infty$, as discussed above $l \sim L k$, so the large $l$ behavior \eqref{kKI:pureAdS} valid for both the polar and axial TLNs computed in the KI formalism becomes ${\cal K}^S_l\to - L k$ and ${\cal K}^V_l\to - L k$.

Using respectively~\eqref{eq:LoveScalarRelation} and~\eqref{eq:LoveVectorRelation}, these KI TLNs translate into the following large $k$ behavior for TLNs derived from the KS formalism:
\be
\lambda_S \; \stackrel{k\to\infty}{\longrightarrow} \; \frac{(L k)^3}{3}\,,
\qquad
\lambda_V \; \stackrel{k\to\infty}{\longrightarrow} \; \frac{(L k)^3}{3}\,.
\ee
This is in precise agreement with Ref.~\cite{Emparan:2017qxd}.

%%%%%%%%%%%%%%%%%%%%%%%%%%%
%%%%%%%%%%%%%%%%%%%%%%%%%%%

\bibliographystyle{JHEP.bst}
\bibliography{TLNs_of_SAdS_arxiv_v1}

%%%%%%%%%%%%%%%%%%%%%%%%%%%
%%%%%%%%%%%%%%%%%%%%%%%%%%%

\end{document}